\documentstyle[12pt]{article}
\begin{document}
\title{Perturbative Renormalization with Flow Equations in Minkowski
Space}

\author{ Georg Keller \\
Rotherschl\"ossli\\
CH-6022 Grosswangen
\and Christoph Kopper, Clemens Schophaus \\
Institut f\"ur Theoretische Physik\\
Universit\"at G\"ottingen\\
Bunsenstra{\ss}e 9\\
D-37073 G\"ottingen}
\maketitle
\thispagestyle{empty}
\vskip 3cm
\begin{abstract}
\noindent We show within the Wilson renormalization group framework how the
flow equation method can be used to prove the perturbative
renormalizability of relativistic massive $\,\Phi_{\,4}^{\,4}\,$.
Furthermore we prove the regularity of the renormalized relativistic
one-particle irreducible n-point Green functions in the region
predicted by axiomatic quantum field theory which ensures that
physical renormalization conditions for the two-point function can be
imposed.
\end{abstract}
\newpage
\setcounter{page}1
\section{Introduction}
The theory of the renormalization group and of effective Lagrangians
which was invented by Wilson and his collaborators in 1974 \cite{wil} has 
proved
to be a rich and powerful method for many branches of quantum field
theory and statistical mechanics. Adapting the Wilson flow equations
to perturbation theory Polchinski first applied this method to the
renormalization problem of perturbative field theory \cite{pol}. Instead of
analysing any complicated divergence/convergence properties of the
general bare or renormalized Feynman diagram, this access solves the
problem of perturbative renormalizability by bounding the solutions
of the system of the first order differential flow equations.

In this paper we continue the programme of two of the authors 
to give mathematically strict proofs
of the perturbative renormalizability of any (by naive power
counting) renormalizable theory of physical interest
using an improved version of Polchinski's method. 
Namely we
show how the flow equation method can be extended to relativistic
theories. The first paper in this series modified and
improved Polchinski's proof of perturbative renormalizability of 
Euclidean massive $\,\Phi_{\,4}^{\,4}\,$ \cite{phi}. Within the Euclidean
framework this improved version of the flow equation method was then applied 
to a
vast range of renormalization problems of perturbative field theory,
e.g. the renormalization of composite operators \cite{com}, the Zimmermann 
identities
\cite{com}, 
the existence of the short distance expansion \cite{sho}, Symanzik's 
improvement programme \cite{sym}, the
construction of the analytical minimal subtraction scheme \cite{anl}, local 
Borel summability for
massive $\,\Phi_{\,4}^{\,4}\,$ \cite{bor} and the renormalization of 
massless
$\,\Phi_{\,4}^{\,4}\,$ \cite{mas} and QED \cite{qem},\cite{qed}.

In order to treat the renormalization problem for relativistic
theories one has to deal with the fact that in momentum space n-point
Green functions can in general only be interpreted as tempered
distributions. Therefore renormalization conditions, i.e. the
requirement that certain n-point Green functions and some of their 
derivatives
take special values at given points in momentum space which ensures
that we are dealing with the right physical constants in the
renormalized theory, can only be imposed if supplementary
regularity properties can be verified. Restricting to a massive scalar field 
theory one has in
particular to ensure that the renormalized two-point Green function
has a pole with residue 1 on the physical mass shell, i.e. the
physical mass shell should be in a region of regularity of the
renormalized amputated two-point Green function. 

In the literature
we found two different ways to handle this problem. The
first way uses renormalization methods
that directly lead to the right renormalization conditions for the
renormalized two-point Green function in every order of perturbation
theory. Furthermore it is shown that in every order the renormalized
n-point Green functions fulfill the LSZ axioms and therefore have the
domain of analyticity predicted by axiomatic quantum field
theory \cite{axf}. Steinmann's method for the renormalization of generalized
retarded n-point functions \cite{sti} and the renormalization method of 
Epstein
and Glaser for time-ordered operator products \cite{ep1},\cite{ep2} solve the 
problem
using this strategy. The other possibility is to renormalize first
using some unphysical renormalization scheme like minimal subtraction
with analytic or dimensional regularization or to subtract at a point
where regularity is evident, e.g. at 0 momentum in a massive theory,
and to ensure afterwards that the required conditions can 
be satisfied in every order of perturbation theory with the help of finite 
renormalizations \cite{hp1}. Therefore it is necessary to show that the
renormalized n-point functions obtained through the first step have 
appropriate regions of regularity for the finite renormalizations to be
well-defined. The singularity regions of unrenormalized Feynman
integrals had been studied in the context of analytical properties of
scattering functions in S-matrix theories \cite{ans}. Because at that time no
method was known that rigorously solved the problems of renormalization,
it had to be taken for granted and was widely accepted that
renormalization did not change the regularity statements.
In 1966 Hepp used the Bogoliubov-Parasiuk subtraction method, which
corresponds to imposing renormalization conditions at 0 momentum, to
prove the perturbative renormalizability of a relativistic massive
scalar field theory \cite{hp2}. Furthermore he showed that for relativistic
massive $\,\Phi_{\,4}^{\,4}\,$ physical renormalization conditions
could be obtained by a finite renormalization \cite{hp2},\cite{hp3}. He only
made a short comment (to our knowledge) on the possibility of
transferring the results about
singularity surfaces of unrenormalized Feynman integrals to the case
of renormalized ones. Chandler proved this to be true for
analytically renormalized Feynman integrals in 1970 \cite{cha}. Finally 
Rivasseau
pointed out to us the strategy of proving for renormalization in parametric
space that the regularity region of the
renormalized two-point Green function is sufficiently large to ensure
physical renormalization conditions on the mass shell \cite{riv}.

In this paper we also start from renormalization conditions at 0
momentum similarly as Hepp did and prove the perturbative
renormalizability of relativistic massive $\,\Phi_{\,4}^{\,4}\,$.
We impose arbitrary conditions at 0 momentum and
obtain renormalized n-point functions. (In fact we fix particularly
simple renormalization conditions at 0 momentum for simplicity of notation,
but the generalization to arbitrary renormalization conditions at 0
momentum can be carried out without difficulties.) Then
we show that these renormalized n-point functions have
regularity regions that admit physical renormalization conditions, 
i.e. the renormalization conditions at 0 momentum can be
chosen in such a way that physical renormalization conditions on the mass
shell are satisfied. In order not to get bothered with the poles
of the amputated connected n-point Green functions for partial sums of 
external 
momenta lying on the mass shell we analyse
the one-particle irreducible Green functions.

The paper is organized as follows: We first derive the flow
equations for one-particle irreducible Green functions in the
Euclidean theory. Then we show certain analyticity properties of the
renormalized Euclidean one-particle irreducible n-point 
functions for complex momenta. We define the relativistic theory with an
$\,\varepsilon$-regularization due to Speer \cite{spe} and prove 
perturbative renormalizability and the fact that the renormalized
one-particle irreducible n-point Green functions become Lorentz
invariant tempered distributions in the limit
$\,\varepsilon\rightarrow 0\,$. Using the analyticity properties of
the renormalized Euclidean theory we then derive the regularity of
the renormalized relativistic one-particle irreducible n-point Green
functions in a region that admits physical renormalization conditions
for the two-point function.

\section{The Flow Equation for regularized Euclidean $\quad\mbox{  
}\quad $ massive $\,\Phi_{\,4}^{\,4}\,$}
In this section we shortly introduce some basic tools of the flow equation
method (for details see \cite{phi},\cite{com}), which are necessary for the 
subsequent considerations.

We use the following regularized free Euclidean propagator 
\begin{equation}
\tilde C_{\alpha}^{\,\alpha_{0}}(p):=
\int\limits_{\alpha_{0}}^{\alpha}
d\alpha' \,e^{\,-\alpha'(\,p_{0}^{2}+\underline{p}^{2}+m^{2})}\quad ,\quad
0<\alpha_{0}\leq\alpha<\infty\quad .
\end{equation} 
Note that this regularization differs from that used in
\cite{phi},\cite{com}, as it respects analyticity in momentum space.
The Fourier transform is denoted as $\,C_{\alpha}^{\,\alpha_{0}}
(x-y)\,$.
The functional Laplace operator $\,\Delta (\alpha,\alpha_{0})\,$ is defined 
as
\begin{displaymath}
\Delta (\alpha,\alpha_{0}):=\frac{1}{2}\int d^{\,4}x\int
d^{\,4}y\ C_{\alpha}^{\,\alpha_{0}}(x-y)\,
\frac{\delta}{\delta\,\Phi(x)}\,\frac{\delta}{\delta\,\Phi(y)}\quad .
\end{displaymath} 
$\,\Phi(x)\,$ may be viewed as an element in $\,S({\rm\bf R}^{4})\,$.
The interaction Lagrangian at scale $\,\alpha_{0}\,$ is given as a formal
power series:
\begin{displaymath}
L^{\,\alpha_{0},\,\alpha_{0}}(\Phi):=\sum_{r\geq 1}g^{\,r}\,
L_{r}^{\,\alpha_{0},\,\alpha_{0}}(\Phi)\quad .
\end{displaymath}
This is the standard Lagrangian including counterterms:
\begin{equation}\label{la0}
L_{r}^{\,\alpha_{0},\,\alpha_{0}}(\Phi):=
\int d^{\,4}x
\ \left (a_{r}^{\,\alpha_{0}}\Phi^{\,2}(x)-b_{r}^{\,\alpha_{0}}
 \Phi(x)\Box\Phi(x)+c_{r}^{\,\alpha_{0}}\Phi^{\,4}(x)\right )\quad,
\end{equation}
where $\,\Box\,$ denotes the 4-dim Laplace operator.

The effective Lagrangian 
\begin{displaymath}
L^{\,\alpha,\,\alpha_{0}}(\Phi):=\sum_{r\geq 1}g^{\,r}\,
L_{r}^{\,\alpha,\,\alpha_{0}}(\Phi)
\end{displaymath}
is defined through
\begin{equation}\label{dla}
e^{\
-\,L^{\,\alpha,\,\alpha_{0}}(\Phi)\ -\,I^{\,\alpha,\,\alpha_{0}}}:=\ 
e^{\,\Delta(\alpha,\alpha_{0})}\
e^{\,-\,L^{\alpha_{0},\,\alpha_{0}}(\Phi)}\quad ,
\end{equation}

where $\,I^{\,\alpha,\,\alpha_{0}}\,$ collects the terms, which satisfy
$\,\frac{\delta}{\delta\,\Phi}\,I^{\,\alpha,\,\alpha_{0}}\equiv 0\,$. 
(Note that as long as these terms appear we have to keep the volume in
(\ref{la0}) finite to be mathematically strict. But as we are only interested
in the $\,\Phi $-dependent terms, we will ignore this point.)

The flow equation for the effective Lagrangian can be obtained by 
differentiation of (\ref{dla}) with respect to $\,\alpha\,$ and 
is given by

\begin{equation}
\partial_{\alpha}\,L^{\,\alpha,\,\alpha_{0}}(\Phi)\, +\, 
\partial_{\alpha}\,I^{\,\alpha,\,\alpha_{0}}\ =\ 
\left[\,\partial_{\alpha}\Delta(\alpha,\alpha_{0})\,\right]\,
L^{\,\alpha,\,\alpha_{0}}(\Phi)
\end{equation}

\begin{displaymath}
-\frac{1}{2}\int\,d^{\,4}x\int\,d^{\,4}y
\,\left (\frac{\delta}{\delta\,\Phi(x)}\,L^{\,\alpha,\,\alpha_{0}}(\Phi
)\right)\left(\partial_{\alpha}\,C_{\alpha}^{\,\alpha_{0}}(x-y)\right )
\frac{\delta}{\delta\,\Phi(y)}\,L^{\,\alpha,\,\alpha_{0}}(\Phi)\quad.
\end{displaymath}

Regarding the fields $\,\Phi\,$ as functions on momentum space, 
$\,L_{r}^{\,\alpha,\,\alpha_{0}}(\Phi)\,$ can be written as

\begin{equation}
L_{r}^{\,\alpha,\,\alpha_{0}}(\Phi)=\sum_{n\geq
2}\ \int\ \prod_{k=1}^{n-1}\,\frac{d^{\,4}p_{k}}{(2\pi)^{\,4}}\ 
\Phi(p_{k})\,\Phi\bigl(-\sum_{j=1}^{n-1}p_{j}\bigr)\ {\cal
L}_{r,n}^{\,\alpha,\,\alpha_{0}}(p_{1},\ldots,p_{n-1})\quad .
\end{equation}
$\,{\cal L}_{r,n}^{\,\alpha,\,\alpha_{0}}\,$ is the $\,r^{{\rm th}}\,$ 
order contribution  to the connected amputated $\,n$-point Green function.

It enjoys the following properties:

a)   $\,{\cal L}_{r,n}^{\,\alpha,\,\alpha_{0}}\,$ may be assumed symmetric
under permutations of $\,p_{1},\ldots,p_{n}:=-\sum_{j=1}^{n-1}p_{j}\,$ .

b)  $\,{\cal L}_{r,n}^{\,\alpha,\,\alpha_{0}}\equiv 0\mbox{  for } n>2r+2\ $ 
(connectedness),

 $\ \,{\cal L}_{r,2k+1}^{\,\alpha,\,\alpha_{0}}\equiv 0\ $ (due to the 
symmetry $\,\Phi\rightarrow -\Phi\,$).

c) $\,{\cal L}_{r,n}^{\,\alpha,\,\alpha_{0}}\,$ is invariant under $\,
O(4)$-transformations of the $\,p_{j}\,$ .

d) $\,{\cal L}_{r,n}^{\,\alpha,\,\alpha_{0}}\,$ is in $\,C^{\,\infty}
(\,[\alpha_{0},
\infty)\times {\rm\bf R}^{\,4(n-1)}\,)\,$ as a function of $\,\alpha\,$ and
$\,p_{1},\ldots,p_{n-1}\,$ .

\section{Flow Equations for one-particle irreducible Green functions}
\subsection{The Generating Functional 
$\,\Gamma^{\,\alpha,\,\alpha_{0}}(\Phi_{c})\,$}

The generating functional $\,W_{c}^{\,\alpha,\,\alpha_{0}}(J)\,$ of the 
pertubative, regularized connected Green functions is given by
\begin{equation}\label{wcj}
W_{c}^{\,\alpha,\,\alpha_{0}}(J):=L^{\,\alpha,\,\alpha_{0}}(\Phi)\,\vert
_{\,\Phi=\tilde C_{\alpha}^{\,\alpha_{0}}\!J}+I^{\,\alpha,\,
\alpha_{0}}-\frac{1}{2}<J,\tilde C_{\alpha}^{\,\alpha_{0}}J>\quad ,
\end{equation}

where 
$\,<f,g>:=\int \frac{d^{\,4}p}{(2\pi)^{\,4}}\,f(-p)g(p)\ $ , 
$\ J\in S({\rm\bf R}^{4})\,$ .
\vskip 0.3cm
The generating functional 
$\,\Gamma^{\,\alpha,\,\alpha_{0}}(\Phi_{c})\,$ 
of the corresponding one-particle irreducible Green functions then can 
be obtained by a Legendre transformation:

Let $\,\delta_{J(p)}:=\delta/\delta J(p)\,$ and 
$\,\frac{1}{(2\pi)^{\,4}}\,\Phi_{c}(-p):=\delta_{J(p)}\,
W_{c}^{\,\alpha,\,\alpha_{0}}(J)\ $ .
Then $\,\Gamma^{\,\alpha,\,\alpha_{0}}(\Phi_{c})\,$ is defined by
\begin{equation}\label{gam}
 \Gamma^{\,\alpha,\,\alpha_{0}}(\Phi_{c}):=
\Bigl [ \,W_{c}^{\,\alpha,\,\alpha_{0}}(J) - < J ,\Phi_{c}>\,   
\Bigr ]_{\,J=J(\Phi_{c})} \quad,
\end{equation}
implying
\begin{equation}\label{help}
\delta_{\Phi_{c}(-p)}\,\Gamma^{\,\alpha,\,\alpha_{0}}
(\Phi_{c})\ =\ \frac{-1}{(2\pi)^{\,4}}\ J(p)\quad.
\end{equation}
In order to compute $\,\Gamma^{\,\alpha,\,\alpha_{0}}(\Phi_{c})\,$ we have to
invert the equation
\begin{equation}\label{phc}
\Phi_{c}(p,J)=\tilde C_{\alpha}^{\,\alpha_{0}}(p)\,\Bigl \{\,(2\pi)^{\,4}
\,\delta_{\Phi
(-p)}\,L^{\,\alpha,\,\alpha_{0}}(\Phi)\,\vert_{\,\Phi=\tilde C_{\alpha}
^{\,\alpha_{0}}\!J}\,-J(p)\,\Bigr \}
\end{equation}
to get $\,J(p,\Phi_{c})\,$. Because $\,L^{\,\alpha,\,\alpha_{0}}(\Phi)
=\sum_{r=1}
^{\infty}\,g^{\,r}L_{r}^{\,\alpha,\,\alpha_{0}}(\Phi)\,$ is a formal
power series in $\,g\,$ and $\,L_{r}^{\,\alpha,\,\alpha_{0}}(\Phi)\,$ is an 
(even due to the symmetry $\,\Phi\rightarrow -\Phi\,$) polynomial
in $\,\Phi\,$ of degree $\leq 2r +2\,$ (connectedness), we can invert 
(\ref{phc}) up to 
any order $\,r\,$ and therefore $\,\Gamma^{\,\alpha,\,\alpha_{0}}(\Phi_{c})\,$ 
is well 
defined in the sense of a formal power series in $\,g\,$. 

We have to keep in mind
that now $\,\Phi_{c}(p)\,$ is viewed as the independent variable (as a 
function $\in 
S( {\rm\bf R^{4}})$) and that $\,J(p,\Phi_{c})\,$ 
is a formal power series in
$\,g\,$, which depends on $\,\alpha\,$ and $\,\alpha_{0}\,$. 
\subsection{The Flow Equation for 
$\,\Gamma^{\,\alpha,\,\alpha_{0}}(\Phi_{c})\,$}
By taking the derivative of (\ref{gam}) with respect to $\,\alpha\,$ we get 
\begin{equation}
\partial_{\alpha}\,\Gamma^{\,\alpha,\,\alpha_{0}}(\Phi_{c})\ = \ 
\left(\,\partial_{\alpha}\,W^{\,\alpha,\,\alpha_{0}}\right )(J)\,
\vert_{\,J=J(\Phi_{c})}
\quad .
\end{equation}
We insert (\ref{wcj}) and obtain
\begin{equation}\label{fg1}
\partial_{\alpha}\,\Gamma^{\,\alpha,\,\alpha_{0}}(\Phi_{c})=\Bigl [
\,\partial_{\alpha}\,\left(\,L^{\,\alpha,\,\alpha_{0}}(\Phi)+I^{\,\alpha,
\,\alpha_{0}}\right )\Bigr ]_{\,\Phi=\tilde C_{\alpha}^{\,\alpha_{0}}
\!J(\Phi_{c})}
\end{equation}
\begin{displaymath}
+\left [\int d^{\,4}q\,\left(\partial_{\alpha}\,\tilde C_{\alpha}
^{\,\alpha_{0}}(q)\right )J(q)\ \delta_{\Phi(q)}\,L^{\,\alpha,\,\alpha_{0}}
(\Phi)-\frac{1}{2}<J,(\partial_{\alpha}\,\tilde C_{\alpha}^{\,\alpha_{0}}
 )J>\right ]_{\,J=J(\Phi_{c})}\quad.
\end{displaymath}

Now we use the flow
equation for $\,L^{\,\alpha,\,\alpha_{0}}(\Phi)\,$:
\begin{equation}\label{fel}
\partial_{\alpha}\,\left(\,L^{\,\alpha,\,\alpha_{0}}(\Phi)+I^{\,\alpha,
\,\alpha_{0}}\right )=[\partial_{\alpha}\,\tilde \Delta(\alpha,\alpha_{0})
\,]\ L^{\,\alpha,\,\alpha_{0}}(\Phi)
\end{equation}
\begin{displaymath}
-\frac{1}{2}<(2\pi)^{\,4}
\delta_{\Phi}\,
L^{\,\alpha,\,\alpha_{0}}(\Phi),(\partial_{\alpha}\,\tilde C_{\alpha}
^{\,\alpha_{0}})\,(2\pi)^{\,4}\delta_{\Phi}\,L^{\,\alpha,\,\alpha_{0}}
(\Phi)>\quad,
\end{displaymath}
\vskip 0.3cm
and get from (\ref{fg1}), (\ref{fel}): 
\begin{equation}
\partial_{\alpha}\,\Gamma^{\,\alpha,\,\alpha_{0}}(\Phi_{c})\ =\
[\partial_{\alpha}\,\tilde\Delta(\alpha,\alpha_{0})\,]\ L^{\,\alpha,
\,\alpha_{0}}(\Phi)\,\vert_{\,\Phi=\tilde C_{\alpha}^{\,\alpha_{0}}\!
J(\Phi_{c})}
\end{equation}
\begin{displaymath}
-\frac{1}{2}<(2\pi)^{\,4}\delta_{\Phi}\,L^{\,\alpha,\,\alpha_{0}}
(\Phi)-J(\Phi_{c}), (\partial_{\alpha}\,\tilde C_{\alpha}^{\,\alpha_{0}}
 )\left((2\pi)^{\,4}\delta_{\Phi}\,L^{\,\alpha,\,\alpha_{0}}(\Phi)
-J(\Phi_{c})
\right )>\,\vert_{\,\Phi=\tilde C_{\alpha}^{\,\alpha_{0}}\!J(\Phi_{c})}
\quad.
\end{displaymath}
\vskip 0.3cm
Together with (\ref{phc}) this yields 
\begin{equation}\label{fg2}
\partial_{\alpha}\,(\,\Gamma^{\,\alpha,\,\alpha_{0}}(\Phi_{c})-
\frac{1}{2}<\Phi_{c},\{\tilde C_{\alpha}^{\,\alpha_{0}}\}^{\,-1}\Phi_{c}>
 )=[\partial_{\alpha}\,\tilde\Delta(\alpha,\alpha_{0})\,]
\ L^{\,\alpha,\,\alpha_{0}}(\Phi)\,\vert_{\,\Phi=\tilde C_{\alpha}
^{\,\alpha_{0}}\!J(\Phi_{c})}\quad .
\end{equation}

In order to arrive at a differential flow equation for $\,\Gamma^{\,\alpha,
\,\alpha_{0}}(\Phi_{c})\,$ we have to express the functional on the 
right hand side of the equation (\ref{fg2}) in terms of 
$\,\Gamma^{\,\alpha,\,\alpha_{0}}(\Phi_{c})\,$. 

We insert (\ref{phc}) into the relation
\begin{equation}\label{help1}
\delta(p+q)\ =\ \delta_{\Phi_{c}(p)}\,\Phi_{c}(-q)
\end{equation}
and with the help of (\ref{help}) we obtain

\begin{equation}\label{help2}
\delta(p+q)=-(2\pi)^{\,8}\int d^{\,4}q'\ 
\Bigl [\tilde C_{\alpha}^{\,\alpha_{0}}
(q)\tilde C_{\alpha}^{\,\alpha_{0}}(q')\ 
\delta_{\Phi_{c}(p)}\delta_{\Phi_{c}(q')}\,
\Gamma^{\,\alpha,\,\alpha_{0}}(\Phi_{c})
\end{equation}
\begin{displaymath}
\delta_{\Phi(-q')}\delta_{\Phi
(q)}\,L^{\,\alpha,\,\alpha_{0}}(\Phi)\,\Bigr ]_{\,\Phi=\tilde C_{\alpha}^
{\,\alpha_{0}}\!J(\Phi_{c})}
+(2\pi)^{\,4}\,\tilde C_{\alpha}^{\,\alpha_{0}}(q)\,\delta_{\Phi_{c}(p)}
\delta_{\Phi_{c}(q)}\,\Gamma^{\,\alpha,\,\alpha_{0}}(\Phi_{c})\quad .
\end{displaymath}
The right hand side of this equation is a formal power series in $\,g\,$. 
(\ref{help2}) has to be 
fulfilled up to any order $\,r\,$ and therefore we get the equations:

\begin{equation}\label{re0}
\delta(p+q)=(2\pi)^{\,4} \,\tilde C_{\alpha}^{\,\alpha_{0}}(q)\
\delta_{\Phi_{c}(p)}\delta_{\Phi_{c}(q)}\,\Gamma_{0}^{\,\alpha,\,\alpha_{0}}
(\Phi_{c})\qquad\mbox{for }\ r=0\quad ,
\end{equation}

\begin{equation}\label{re1}
\sum_{k=0}^{r-1}-(2\pi)^{\,4}\int d^{\,4}q'\ \tilde C_{\alpha}^{\,\alpha_{0}}
(q)\tilde C_{\alpha}^{\,\alpha_{0}}(q')\ \hat\Gamma^{\,\alpha,\,\alpha_{0}}
_{r-k}(q,-q',\Phi_{c})\ \delta_{\Phi_{c}(p)}\delta_{\Phi_{c}(q')}\,
\Gamma_{k}^{\,\alpha,\,\alpha_{0}}(\Phi_{c})
\end{equation}
\begin{displaymath}
+\,(2\pi)^{\,4}\,\tilde 
C_{\alpha}^{\,\alpha_{0}}(q)\ \delta_{\Phi_{c}(p)}\delta_{\Phi_{c}(q)}
\,\Gamma_{r}^{\,\alpha,\,\alpha_{0}}(\Phi_{c})=0
\qquad,\ r>0\quad,
\end{displaymath}
with the definition
\begin{equation}\label{hga}
\hat\Gamma_{r}^{\,\alpha,\,\alpha_{0}}(q,p,\Phi_{c})\ :=\ 
(2\pi)^{\,4}\left \{\,\delta_{\Phi(p)}\delta_{\Phi(q)}\,L^{\,\alpha,\,
\alpha_{0}}(\Phi)\,\vert_{\,\Phi=\tilde C_{\alpha}^{\,\alpha_{0}}\!J(
\Phi_{c})}\right \}_{r}\quad .
\end{equation}
We now can insert (\ref{re0}) into (\ref{re1}) and obtain
\begin{equation}\label{re2}
\hat \Gamma_{r}^{\,\alpha,\,\alpha_{0}}(q,p,\Phi_{c})=(2\pi)^{\,4}\,
\delta_{\Phi_{c}(p)}\delta_{\Phi_{c}(q)}\,\Gamma_{r}^{\,\alpha,\,\alpha_{0}}
(\Phi_{c})
\end{equation}
\begin{displaymath}
-\,(2\pi)^{\,4}\sum_{k=1}^{r-1}\int d^{\,4}q'\ \tilde C_{\alpha}^{\,
\alpha_{0}}(q')\ \hat\Gamma_{r-k}^{\,
\alpha,\,\alpha_{0}}(q,-q',\Phi_{c})\ \delta_{\Phi_{c}(p)}\delta_{\Phi_{c}
(q')}\,\Gamma_{k}^{\,\alpha,\,\alpha_{0}}(\Phi_{c})\quad .
\end{displaymath}

(\ref{re2}) is a recursive relation for 
$\,\hat\Gamma_{r}^{\,\alpha,\,\alpha_
{0}}(q,p,\Phi_{c})\,$ and allows us to express $\,\hat\Gamma_{r}^{\,
\alpha,\,\alpha_{0}}(q,p,\Phi_{c})\,$ in terms of $\,\Gamma_{k}^{\,\alpha,
\,\alpha_{0}}(\Phi_{c})\ ,\ k=1,\ldots,r\,$ .

We compare the left and the right hand side of (\ref{fg2}) in powers of 
$\,g\,$,
use the definition (\ref{hga}) and end up with a differential flow equation 
for $\,\Gamma_{r}^{\,\alpha,\,\alpha_{0}}(\Phi_{c})\,$:
\begin{equation}\label{fg3}
\partial_{\alpha}\,\Gamma_{r}^{\,\alpha,\,\alpha_{0}}(\Phi_{c})\ =\ 
\frac{1}{2}\int d^{\,4}p\ (\partial_{\alpha}\,\tilde C_{\alpha}^{\,\alpha_{0}}
(p))
\ \hat\Gamma_{r}^{\,\alpha,\,\alpha_{0}}(p,-p,\Phi_{c})\quad ,\quad
r\geq1\ .
\end{equation}
\subsection{One-particle irreducible Green functions}
$\,\Gamma^{\,\alpha,\,\alpha_{0}}_{r}(\Phi_{c})\,$ is an even polynomial 
in $\,\Phi_{c}\,$:
\begin{equation}\label{gfg}
\Gamma_{r}^{\,\alpha,\,\alpha_{0}}(\Phi_{c})=\sum_{n\geq 2}\,\int\,\prod_
{k=1}^{n-1}\,\frac{d^{\,4}p_{k}}{(2\pi)^{\,4}}\ \Phi_{c}(p_{k})\,\Phi_{c}
\bigl (-\sum_{j=1}^{n-1}p_{j}\bigr )\ 
\Gamma_{r,n}^{\,\alpha,\,\alpha_{0}}(p_{1},\ldots,p_{n-1})\quad.
\end{equation}
$\,\Gamma_{r,n}^{\,\alpha,\,\alpha_{0}}\,$ is the momentum space
regularized one-particle irreducible $\,n$-point Green function of
order $\,r\,$.

It is defined to be symmetric under permutations of $\,p_{1},\ldots,p_{n}:=
-\sum_{j=1}^{n-1}p_{j}\,$, it is invariant under $\,O(4)$-transformations 
of the $\,p_{j}\,$ and it is
in $\,C^{\,\infty}(\,[\alpha_{0},\infty)\times {\rm\bf R}^{\,4(n-1)}
\,)\,$ as a function of $\,\alpha\,$ and $\,p_{1},\ldots,p_{n-1}\,$ .
We have: $\ \Gamma_{r,n}^{\,\alpha,\,\alpha_{0}}
\equiv 0\ \mbox{ for }\ n>2r+2\ $ (connectedness).
(\ref{fg3}) rewritten for the coefficient functions $\,\Gamma_{r,n}^{\,
\alpha,\,\alpha_{0}}(p_{1},\ldots,p_{n-1})\,$ yields:
\begin{equation}\label{fg4}
\partial_{\alpha}\,\Gamma_{r,n}^{\,\alpha,\,\alpha_{0}}(p_{1},\ldots,
p_{n-1})=\frac{1}{2}\int\frac{d^{\,4}p}{(2\pi)^{\,4}}\ (\partial_{\alpha}\,
\tilde
C_{\alpha}^{\,\alpha_{0}}(p))\ \hat \Gamma_{r,n+2}^{\,\alpha,\,
\alpha_{0}}(p,-p,p_{1},\ldots,p_{n-1})\quad,
\end{equation}
where
\begin{equation}\label{ket}
\hat\Gamma_{r,n+2}^{\,\alpha,\,\alpha_{0}}(p,-p,p_{1},\ldots,p_{n-1}):=
(n+1)(n+2)\,\Gamma_{r,n+2}^{\,\alpha,\,\alpha_{0}}(p,-p,p_{1},\ldots,
p_{n-1})
\end{equation}
\begin{displaymath}
-\sum_{v=2}^{r}\sum_{\{a_{j}\},\{b_{j}\}}(-1)^{v}
K^{\,v}(b_{1},\ldots,b_{v})\Bigl [\,\prod_{k=1}^{v-1}\,\tilde C_{\alpha}^{\,
\alpha_{0}}(q_{k}')\ \Gamma_{a_{k},b_{k}+2}^{\,\alpha,\,\alpha_{0}}
(q_{k-1}',p_{i_{k}+1},\ldots,p_{i_{k}+b_{k}})
\end{displaymath}
\begin{displaymath}
\Gamma_{a_{v},b_{v}+2}^{\,\alpha,\,\alpha_{0}}(q_{v-1}',-p,p_{i_{v}+1},
\ldots,p_{n-1})\,\Bigr ]_{\,{\rm symm.}}\quad,
\end{displaymath}
with
\begin{displaymath}
q_{0}'=p\quad ,\quad q_{k}'=p+\sum_{j=1}^{b_{1}+\ldots+b_{k}}p_{j}\quad
,\quad a_{j}>0\quad ,\quad 
b_{j}=0,2,4,\ldots,n\quad ,\quad i_{k}=\sum_{j=1}^{k-1}b_{j}\quad.
\end{displaymath}
The sum is over all $\,\{a_{j}\}\,$ with $\,\sum_{j=1}^{v}a_{j}=r\,$
and over all $\,\{b_{j}\}\,$ with $\,\sum_{j=1}^{v}b_{j}=n\,$.\\
$\, K^{\,v}(b_{1},\ldots,b_{v})\, $ is a combinatorial factor, which could 
be 
computed with the help of (\ref{re2}) and $\,[\ldots]_{{\rm symm.}}\,$ 
indicates the symmetrization  
operation with respect to $\,p_{1},\ldots,p_{n}=-\sum_{j=1}^{n-1}p_{j}\,$.

We use (\ref{phc}) and (\ref{gam}) to compute the lowest order
contribution in powers of $\,\tilde C_{\alpha}^{\,\alpha_{0}}\,$ to 
$\,\Gamma_{r}^{\,\alpha,\,\alpha_{0}}\,$ and
as $\,\tilde C_{\alpha_{0}}^{\,\alpha_{0}}(p)
\equiv 0\,$, we conclude that for $\,\alpha\rightarrow\alpha_{0}\,$ $\,
{\cal L}_{r,n}^{\,\alpha_{0},\,\alpha_{0}} 
\equiv \Gamma_{r,n}^{\,\alpha_{0},\,\alpha_{0}}\,$ and therefore we get 
from (\ref{la0}) the boundary values at $\,\alpha=\alpha_{0}\,$:
\begin{equation}\label{ra0}
\Gamma_{r,2}^{\,\alpha_{0},\,\alpha_{0}}(p)\ =\ a_{r}^{\,\alpha_{0}}+
b_{r}^{\,\alpha_{0}}p^{2}\quad ,\quad 
\Gamma_{r,4}^{\,\alpha_{0},\,\alpha_{0}}(p_{1},p_{2},p_{3})=c_{r}^{\,
\alpha_{0}}\quad,\quad \Gamma_{r,n}^{\,\alpha_{0},\,\alpha_{0}}(\vec p)
\equiv 0\quad\mbox{for}\ n>4\quad.
\end{equation}
$\,\vec p\,$ denotes the tuple $\,(p_{1},\ldots,p_{n-1})\,$ .

(\ref{ra0}) implies
\begin{equation}\label{wra}
\partial_{p}^{\omega}\,\Gamma_{r,n}^{\,\alpha_{0},\,\alpha_{0}}(\vec
p)\equiv 0\quad\mbox{for}\quad n+\vert\omega\vert>4\quad,
\end{equation}
where $\ \omega=(\omega_{1}^{0},\ldots,\omega_{1}^{3},\omega_{2}^{0},
\ldots,\omega_{n-1}^{3})\ $ is a multiindex 
and 
\begin{center}$\,\partial_{p}^{\omega}\equiv\partial_{p_{1}}^{\omega_{1}^{0}}
\ldots 
\partial_{p_{n-1}}^{\omega_{n-1}^{3}}\qquad$ , $\qquad\vert\omega\vert=
\sum_{\mu=0}^{
3}\sum_{j=1}^{n-1}\omega_{j}^{\mu}\quad $.
\end{center}

With the help of (\ref{fg4}), (\ref{ket}) and (\ref{ra0}) all one-particle 
irreducible 
$\,n$-point Green functions of any order $\,r\,$ can be computed by 
integrating successively (\ref{fg4}) with respect to $\,\alpha\,$
from the lower bound $\,\alpha_{0}\,$ up to the new parameter $\,\alpha\,$, 
following the standard induction
scheme of the flow equation method (see \cite{phi}) upwards in $\,r\,$ and for 
given $\,r\,$ downwards in $\,n\,$.

For illustration we can interpret the contributions to $\,\hat\Gamma_{
r,n+2}^{\,\alpha,\,\alpha_{0}}\,$ in (\ref{ket}) in terms of Feynman graphs:
We suppose $\,\Gamma_{r,n}^{\,\alpha,\alpha_{0}}\,$ can be written as a sum
of one-particle irreducible Feynman graphs where every internal line is a 
function of the parameter $\,\alpha\,$. If we now take the derivative of 
$\,\Gamma_{r,n}^{\,\alpha,\,\alpha_{0}}\,$ with respect to $\,\alpha\,$,
we can divide the expression obtained into contributions of two 
different types. Contributions of the first type stay one-particle 
irreducible, if we remove the line on which the derivative acts. The first
term on the right hand side of (\ref{ket}) can be interpreted as a sum of
all these contributions. Contributions of the second type become one-particle
reducible, if we remove the line on which the derivative acts. The sum of
these contributions corresponds to the second term on the right hand side
of (\ref{ket}).

Due to these considerations we conclude that by integrating
successively (\ref{fg4}) we indeed get the one-particle irreducible 
Green functions (the proof could be carried out by induction in $\,r\,$
and $\,n\,$).

\section{Renormalizability and Analyticity of the $\quad \mbox{  }\qquad 
\mbox{     
     }\qquad$ one-particle irreducible Green functions}  
Let us now change the index pair $\,(r,n)\,$ and take 
\begin{equation}\label{ls}
l:=r-\frac{n}{2}+1\quad\mbox{ and }\quad s:=2r-\frac{n}{2}
\end{equation}
as a new index pair to number our Green functions.(In the language of 
Feynman graphs $\,l\,$ corresponds to the number of loops and $\,s\,$
to the number of internal lines of an (unrenormalized) graph.) For our new 
indices
(\ref{wra}) reads: 
\begin{equation}\label{ra1}
\partial_{p}^{\omega}\,\Gamma_{l,s}^{\,\alpha_{0},\,\alpha_{0}}(\vec p)
\equiv 0\quad\mbox{ for }\quad \frac{\vert\omega\vert}{2}+s > 2l\quad.
\end{equation}
Furthermore we have:
\begin{equation}\label{ra2}
\Gamma_{l,s}^{\,\alpha,\,\alpha_{0}}(\vec p)\equiv 0\qquad\mbox{for}
\quad\underbrace{s<2l-1}_{n<2}\quad,\quad \underbrace{l<0}_{n>2r+2}
\quad,\quad \underbrace{s<0}_{n>4r}\quad.
\end{equation}
The flow equation (\ref{fg4}) written for the new indices is
\begin{equation}\label{fg5}
\partial_{\alpha}\,\Gamma_{l,s}^{\,\alpha,\,\alpha_{0}}(\vec p)=
\frac{1}{2}\int\frac{d^{\,4}p}{(2\pi)^{\,4}}\ (\partial_{\alpha}\,
\tilde C_{\alpha}^{\,\alpha_{0}}(p))\ \hat \Gamma_{l-1,s-1}^{\,\alpha,\,
\alpha_{0}}(p,-p,\vec p)\quad,
\end{equation}
where
\begin{equation}\label{kes}
\hat\Gamma_{l-1,s-1}^{\,\alpha,\,\alpha_{0}}(p,-p,\vec p):=
h_{l}^{s}\ \Gamma_{l-1,s-1}^{\,\alpha,\,\alpha_{0}}(p,-p,\vec p)
\end{equation}
\begin{displaymath}
-\sum_{v=2}^{s-l+1}\sum_{\{c_{j}\},\{d_{j}\}}
(-1)^{v}\,K^{\,v}(b_{1},\ldots,b_{v})
\Bigl [\,\prod_{k=1}^{v-1}\,\tilde C_{\alpha}^{\,
\alpha_{0}}(q_{k}')\ \Gamma_{c_{k},d_{k}}^{\,\alpha,\,\alpha_{0}}
(\vec p_{k})
\ \Gamma_{c_{v},d_{v}}^{\,\alpha,\,\alpha_{0}}(\vec p_{v})\,\Bigr ]_{\,
{\rm symm.}}\quad.
\end{displaymath}
The sum is over all $\,\{c_{j}\}\,,\,\{d_{j}\}\,$ with 
$\,\sum_{j=1}^{v}c_{j}=l-1\,$ and $\,\sum_{j=1}^{v}d_{j}+v=s\,$, where
$\,c_{j},d_{j}\geq 0\,$.

Furthermore we have:
\begin{displaymath}
\vec p_{k}=(q_{k-1}',p_{i_{k}+1},\ldots,p_{i_{k}+b_{k}})\quad ,\quad 
b_{k}=2d_{k}-4c_{k}+2\quad ,\quad i_{k}=\sum_{j=1}^{k-1}b_{j}\quad ,
\end{displaymath}
\begin{displaymath}
q_{0}'=p\quad ,\quad
q_{k}'=p+\sum_{j=1}^{b_{1}+\ldots+b_{k}}p_{j}\quad ,\quad 
\vec
p_{v}=(q_{v-1}',p_{i_{v}+1},\ldots,p_{n-1},-p)\quad,\quad h_{l}^{s}=(n+2)(
n+1)\ .
\end{displaymath}

\subsection{Integral Representation for $\partial_{p}^{\omega}\,
\Gamma_{l,s}^{\,\alpha,\,
\alpha_{0}}(\vec p\,)\,$}
By successive integration of the flow equation (\ref{fg5}) we now want to
derive an integral representation for $\partial_{p}^{\omega}\,\Gamma_{l,s}^{
\,\alpha,\,\alpha_{0}}\,$.
Note that because we are in a massive theory we may
integrate the flow equation (\ref{fg5}) and its momentum derivatives
up to infinity with respect to $\,\alpha\,$ since the mass provides an
exponential infrared cutoff for large $\,\alpha\,$.

We use the following boundary conditions (see also \cite{phi},\cite{com}): 
At 
$\,\alpha=\alpha_{0}\,$ we impose (\ref{ra1}). For the so-called relevant
and marginal terms with $\,\frac{\vert\omega\vert}{2}+s\leq 2l\,$ we impose 
renormalization conditions at $\,\alpha=\infty\,$ by
fixing the values of $\,\Gamma
^{\,\infty,\,\alpha_{0}}_{l,2l}(0)\,$, $\,\Gamma^{\,\infty,\,\alpha_{0}}_{l,
2l-1}(0)\,$ and $\,\partial_{\mu}\partial_{\nu}\,\Gamma^{\,\infty,\,
\alpha_{0}}_{l,2l-1}(0)\,$ (We restrict to momentum 0 because we want
to analyse the corresponding relativistic theory later on, and there
it is convenient to start by renormalizing at 0 momentum). We 
have to distinguish three cases:
\vskip 0.5 cm
1.$\ s>2l$
\begin{equation}\label{sl1}
\Gamma_{l,s}^{\,\alpha,\,\alpha_{0}}(\vec p)=\int\limits_{\alpha_{0}}^{
\alpha}
d\alpha'\ \partial_{\alpha'}\,\Gamma_{l,s}^{\,\alpha',\,\alpha_{0}}(\vec p)=
\int\limits_{\alpha_{0}}^{\alpha}d\alpha'\ \mbox{(r.h.s. of (\ref{fg5}))}
\quad.
\end{equation}

2.$\ s=2l $
\begin{equation}\label{sl2}
\Gamma_{l,s}^{\,\alpha,\,\alpha_{0}}(\vec p)=\Gamma_{l,s}^{\,\alpha,\,
\alpha_{0}}(0)
+\sum_{j=1}^{3}\,\sum_{\mu=0}^{3}\,p_{\mu,j}\int\limits_{0}^{1}d\lambda\,
\left(\,\partial_{\mu,j}\,\Gamma_{l,s}^{\,\alpha,\,\alpha_{0}}\right)
(\lambda\,\vec p)\quad,
\end{equation}
with
\begin{equation}\label{gao}
\Gamma_{l,s}^{\,\alpha,\,\alpha_{0}}(0)=\Gamma_{l,s}^{\,\infty,\,\alpha_{0}}
(0)-
\int\limits_{\alpha}^{\infty}d\alpha'\,\underbrace{\,\partial_{\alpha'}\,
\Gamma_{l,s}^{\,\alpha',
\,\alpha_{0}}(0)\,}_{\mbox{r.h.s. of (\ref{fg5})}}
\end{equation}
and
\begin{displaymath}
\partial_{\mu,j}\,\Gamma_{l,s}^{\,\alpha,\,\alpha_{0}}(\vec p)=
\int\limits_{\alpha_{0}}^{\alpha}d\alpha'\,\partial_{\mu,j}\,\underbrace{\,
\partial_{\alpha'}\,\Gamma_{l,s}^{\,\alpha',\,\alpha_{0}}(\vec p)\,}_
{\mbox{r.h.s. of (\ref{fg5})}}\quad.
\end{displaymath}
 
3.$\ s=2l-1$
\begin{equation}\label{sl3}
\Gamma_{l,s}^{\,\alpha,\,\alpha_{0}}(p)=\underbrace{\Gamma_{l,s}^{\,
\alpha,\,\alpha_{0}}(0)}_{\rm{as\ in\ (\ref{gao}})}
+\frac{1}{2}\sum_{\mu,\nu=0}^{3}\,p_{\mu}\,p_{\nu}\,
\partial_{\mu}\,\partial_{\nu}\,\Gamma_{l,s}^{\,
\alpha,\,\alpha_{0}}(0)
\end{equation}
\begin{displaymath}
+\sum_{\mu,\mu',\,\mu''=0}^{3}\,p_{\mu}\int\limits
_{0}^{1}d\lambda_{1}\,p_{\mu'}\lambda_{1}\int\limits_{0}^{1}d\lambda_{2}
\,p_{\mu''}
\lambda_{2}\lambda_{1}\int\limits_{0}^{1}d\lambda_{3}\left(\partial_{\mu''}
\partial_{\mu'}
\partial_{\mu}\,\Gamma_{l,s}^{\,\alpha,\,\alpha_{0}}\right)(\lambda_{3}
\lambda_{2}\lambda_{1}\,p)\quad,
\end{displaymath}
with
\begin{displaymath}
\partial_{\mu}\,\partial_{\nu}\,\Gamma_{l,s}^{\,\alpha,\,\alpha_{0}}(0)=
\partial_{\mu}\,\partial_{\nu}\,\Gamma_{l,s}^{\,\infty,\,\alpha_{0}}(0)-
\int\limits_{\alpha}^{\infty}d\alpha'\,\partial_{\mu}\,\partial_{\nu}\,
\underbrace{\partial_{\alpha'}\,\Gamma_{l,s}^{\,\alpha',\,\alpha_{0}}(0)\,}
_{\mbox{r.h.s. of (\ref{fg5})}}
\end{displaymath}
and
\begin{displaymath}
\partial_{\mu''}\,\partial_{\mu'}\,\partial_{\mu}\,\Gamma_{l,s}^{\,\alpha,\,
\alpha_{0}}
(p)=\int\limits_{\alpha_{0}}^{\alpha}d\alpha'\,\,\partial_{\mu''}\,
\partial_{\mu'}\,\partial_{\mu}\,\underbrace{\partial_{\alpha'}\,
\Gamma_{l,s}^{\,\alpha',
\,\alpha_{0}}(p)\,}_{\mbox{r.h.s. of (\ref{fg5})}}\quad.
\end{displaymath}

For $\,l=0\,$ we get from (\ref{ra2}), (\ref{fg5}), (\ref{sl1}) and 
(\ref{sl2}):
\begin{equation}\label{sta}
\Gamma_{0,0}^{\,\alpha,\,\alpha_{0}}(\vec
p)=\Gamma_{0,0}^{\,\infty,\,\alpha_{0}}(0)\quad\mbox{and}\quad
\Gamma_{0,s}^{\,\alpha,\,\alpha_{0}}(\vec p)\equiv
0\quad\mbox{for}\ s>0\quad.
\end{equation}
We impose for reasons of simplicity the 
renormalization conditions:
\begin{equation}\label{ren}
\Gamma_{0,0}^{\,\infty,\,\alpha_{0}}(0):=c_{1}^{\,R}\quad\mbox{and}
\quad
\Gamma_{l,2l}^{\,\infty,\,\alpha_{0}}(0)=0\,,\,\Gamma_{l,2l-1}^{\,\infty,\,
\alpha_{0}}(0)=0\,,\,\partial_{\mu}\,\partial_{\nu}\,\Gamma_{l,2l-1}^
{\,\infty,
\,\alpha_{0}}(0)=0\quad\mbox{for}\quad l>0\ .
\end{equation}
Note that once we have fixed the renormalization conditions 
the bare parameters appearing in
(\ref{ra0}) are determined uniquely \cite{phi},\cite{com}.

Using (\ref{fg5}), (\ref{sl1}), (\ref{sl2}), (\ref{sl3}) and the starting
point (\ref{sta}), we obtain by induction an integral representation for
$\,\partial_{p}^{\omega}\,\Gamma_{l,s}^{\,\alpha,\,\alpha_{0}}\,$:

{\bf Lemma 1.}
\begin{equation}\label{int}
\partial_{p}^{\omega}\,\Gamma^{\,\alpha,\,\alpha_{0}}_{l,s}(\vec
p)=\int\limits_{0}^{1}d\lambda_{1}\ldots\int\limits_{0}^{1}
d\lambda_{\sigma(l,s)}
\int\limits_{\alpha_{0}}^{\infty}d\alpha_{1}\ldots\int
\limits_{\alpha_{0}}^{\infty}d\alpha_{s}\
\partial_{p}^{\omega}\,G^{\,\alpha}_{l,s}(\vec \alpha,\vec \lambda,\vec p)
\quad.
\end{equation}
{\it $\,\vec\alpha\,$ and $\,\vec\lambda\,$ denote the tuples $\,(\alpha_{1},
\ldots,\alpha_{s})\,$ and $\,(\lambda_{1},\ldots,\lambda_{\sigma(l,s)})\,$
and $\,\partial_{p}^{\omega}\,G_{l,s}^{\,\alpha}\,$ obeys the bounds}
\begin{equation}\label{bon}
\vert\partial_{p}^{\omega}\,G^{\,\alpha}_{l,s}(\vec\alpha,\vec\lambda,
\vec p)\vert\ \leq\ e^{\,-m^{2}\sum_{j=1}^{s}\alpha_{j}}\,P
(\vert\vec p\vert)\,Q(\sqrt\alpha_{1},\ldots,\sqrt\alpha_{s})\quad,
\end{equation}
{\it where $\,P\,$ is a polynomial with nonnegative coefficients
-- independent of $\,\alpha\,$ -- \\in $\,\vert p_{0,1}\vert,\ldots,
\vert p_{3,n-1}\vert\,$ and $\,Q\,$ is a nonnegative rational function which 
has no poles for $\,\alpha_{i}>0\, $.}  

{\it Proof.} The induction scheme is quite simple as on the right hand side
of (\ref{fg5}) there only appear contributions up to $\,(l-1,s-1)\,$ and 
therefore the induction could proceed for example in $\,l+s\,$.
To give a short indication how the elementary proof works 
let us carry out the induction step for 
$\,s>2l\ , \vert\omega\vert=0\,$:

We employ the induction hypothesis 
(\ref{int}) on the right hand side of (\ref{sl1}) and interchange the 
loop-integration $\,d^{\,4}p\,$ with the $\,d\lambda_{1}\ldots
d\lambda_{\sigma_{{\rm max}}}d\alpha_{1}\ldots d\alpha_{s-1}\,$ integration 
which is justified because of (\ref{bon}). This yields (\ref{sgl}), 
(\ref{gtl}). Now we check that the new $\,G_{l,s}^{\,\alpha}\,$ obeys the
bounds (\ref{bon}) and the induction step is completed. The induction step
for $\,\vert\omega\vert\not=0\,$ and $\,s=2l\,$, $\,s=2l-1\,$ is 
analogous. $\,\Box\,$
 
$\,G_{l,s}^{\,\alpha}(\vec \alpha,\vec \lambda,\vec p)\, $ and 
$\,\sigma(l,s)\,$ are 
determined through the following recursive relations:  
\vskip0.3cm
1. $ s>2l $
\begin{equation}\label{sgl}
G^{\,\alpha}_{l,s}(\vec \alpha,\vec \lambda,\vec p)=
\tilde G_{l,s}(\vec \alpha,\vec \lambda',\vec p)\, 
\Theta(\alpha-\alpha_{s})
\end{equation}
with
\begin{equation}\label{gtl}
\tilde G_{l,s}(\vec \alpha,\vec\lambda',\vec p):=\frac{1}{2}
\int\frac{d^{\,4}p}{(2\pi)^{\,4}}\ 
e^{\,-\alpha_{s}(\,p^{2}+m^{2})}\,\Bigl \{h_{l}^{s}\,
G^{\,\alpha_{s}}_{l-1,s-1}(\vec \alpha,\vec \lambda,\vec
p,p,-p) 
\end{equation}
\begin{displaymath}
-
\sum_{v=2}^{s-l+1}\sum_{\{c_{j}\},\{d_{j}\}}(-1)^{v}K^{\,v}
(b_{1},\ldots,b_{v})
\Bigl [\,\prod_{k=1}^{v-1}\,e^{\,-\alpha_{w_{k}}(\,q_{k}'^{2}+m^{2})}
\,\Theta(\alpha_{s}-\alpha_{w_{k}})\,G_{c_{k},d_{k}}^{\,\alpha_{s}}
(\vec \alpha_{k},\vec\lambda_{k},\vec p_{k})
\end{displaymath}
\begin{displaymath}
G_{c_{v},d_{v}}^{\,\alpha_{s}}
(\vec \alpha_{v},\vec\lambda_{v},\vec p_{v})\,\Bigr ]_{\,{\rm
symm.}}\Bigr \}\quad .
\end{displaymath}
\vskip 0.3cm
The sum is over all $\,\{c_{j}\}\,,\,\{d_{j}\}\,$ with 
$\,\sum_{j=1}^{v}c_{j}=l-1\,$ and $\,\sum_{j=1}^{v}d_{j}+v=s\,$, where
$\,c_{j},d_{j}\geq 0\,$.

Furthermore we have:
\begin{displaymath}
\vec\alpha_{k}=(\alpha_{f_{k}+1},\ldots,\alpha_{f_{k}+d_{k}})\quad ,\quad 
\vec\lambda_{k}=(\lambda_{u_{k}+1},\ldots,\lambda_{u_{k}+
\sigma(c_{k},d_{k})})\quad ,\quad\vec\lambda'=(\lambda_{1},\ldots,
\lambda_{\tilde\sigma(l,s)})\quad,
\end{displaymath}
\begin{displaymath}
w_{k}=\sum_{j=1}^{v}d_{j}+k\quad,\quad f_{k}=\sum_{j=1}^{k-1}d_{j}\quad,
\quad u_{k}=\sum_{j=1}^{k-1}\sigma(c_{j},d_{j})\quad.
\end{displaymath}
As we can add to (\ref{int}) as many integrals $\,\int_{0}^{1}d\lambda_{j}\,$
with new variables $\,\lambda_{j}\,$ as we like without changing anything 
we take the maximum number of 
$\,\lambda$-integrals which appear on the right hand side of (\ref{fg5}) 
during one induction step as our new number of $\lambda$-integrals.
For $\,s>2l\,$ we set
\begin{displaymath}
\,\sigma(l,s):=\tilde \sigma(l,s):=
\max_{v}\,\{\sigma(l-1,s-1),n_{v}\}\  \mbox{ with }\  
n_{v}:=\max_{\{c_{j}\},
\{d_{j}\}}\,\bigl\{\sum_{k=1}^{v}\sigma({c_{k},d_{k}})\bigr \}\ .
\end{displaymath}

2. $ s=2l $
\begin{equation}\label{s2l}
G^{\,\alpha}_{l,s}(\vec \alpha,\vec\lambda,\vec p)=
-\tilde G_{l,s}(\vec
\alpha,\vec\lambda',0)\,\Theta(\alpha_{s}-\alpha)+\sum_{j=1}^{3}\sum_{
\mu=0}^{3}p_{\mu,j}
\ \Bigl (\partial_{\mu,j}\,
\tilde G_{l,s}\Bigl )(\vec \alpha,\vec\lambda',
\lambda_{\sigma(l,s)}\,\vec p)\, \Theta(\alpha-\alpha_{s})\quad.
\end{equation}
Here we set $\,\sigma(l,s):=1+\tilde \sigma(l,s)\,$ .
\vskip 0.3cm
3. $ s=2l-1 $
\begin{equation}\label{s1l}
G^{\,\alpha}_{l,s}(\vec \alpha,\vec \lambda,
p)=\Bigl [- \tilde G_{l,s}(\vec \alpha,\vec\lambda',0)
-\frac{1}{2}\sum_{\nu,\,\mu=0}^{3}\,p_{\mu}\,p_{\nu}\,\partial_{\mu}
\partial_{\nu}\tilde G_{l,s}(\vec \alpha,\vec\lambda',0)\Bigr ]\,\Theta
(\alpha_{s}-\alpha)
\end{equation}
\begin{displaymath}
+\sum_{\mu,\,\mu',\,\mu''=0}^{3}\,\lambda_{\sigma(l,s)-2}^{2}\,
\lambda_{\sigma(l,s)-1}\,p_{\mu}\,p_{\mu'}\,p_{\mu''}\,\Bigl (
\partial_{\mu''}\partial_{\mu'}\partial_{\mu}\,\tilde
G_{l,s}\Bigr )(\vec
\alpha,\vec\lambda',\lambda_{\sigma(l,s)-2}\,\lambda_{\sigma(l,s)-1}\,
\lambda_{\sigma(l,s)} \,p)\,\Theta(\alpha-\alpha_{s})
\end{displaymath}    

$\,\sigma(l,s):=3+\tilde \sigma(l,s)\,$ .
\vskip 0.3cm
Note that, as long as we keep $\,\alpha_{0}>0\,$ and the external
momenta are real, we have absolute
convergence of the integrals in (\ref{int}), even for
$\,\alpha=\infty\ $.

\subsection{Convergence of the Integral Representation}
In this section we want to examine the convergence of the integral
representation of $\,\partial_{p}^{\omega}\,\Gamma_{l,s}^{\,\alpha,\,
\alpha_{0}}\,$ as $\,\alpha_{0}\rightarrow 0\,$ and
$\alpha\rightarrow\infty\ $ in a complex domain obtained by continuing 
$\,p_{0,1},p_{0,2},\ldots,p_{0,n-1}\,$ to complex values (see (\ref{img})
below).

From (\ref{sgl}), (\ref{gtl}), (\ref{s2l}) and (\ref{s1l}) we can realize
inductively in $\,l+s\,$ that
$\,\partial_{p}^{\omega}\,G^{\,\alpha}_{l,s}\,$ may be analytically 
continued 
in the zero components of the external momenta into any complex domain:
\begin{equation}\label{kog}
\partial_{p}^{\omega}\,G^{\,\alpha}_{l,s}(\vec
\alpha,\vec\lambda,p_{0,1}+ik_{0,1},\underline{p}_{1},\ldots,p_{0,n-1}
+ik_{0,n-1},\underline{p}_{n-1})
\end{equation}
is well-defined for finite positive $\,\alpha,\vec\alpha\,$ and polynomially 
bounded in $\,\vec p\,$ for \\
$\,k_{0,1},\ldots,k_{0,n-1}\in {\rm\bf R}\,$.
From now on we restrict the imaginary parts $\,\vec k_{0}=(k_{0,1},\ldots,
k_{0,n-1})\,$ to 
\begin{equation}\label{img}
\vert\,\sum_{j\in\tau_{a}}k_{0,j}\,\vert\leq 2(m-\eta)\quad\mbox{ for all }
\quad
\tau_{a}\subseteq\{1,\ldots,n\}\quad,\quad k_{0,n}=-\sum_{j=1}^{n-1}k_{0,j}
\end{equation}
and want to show that we still can control the limit $\,\alpha\rightarrow
\infty\,$
of $\,\partial_{p}^{\omega}\,\Gamma_{l,s}^{\,\alpha,\,\alpha_{0}}\,$. Here
and in the following $\,\eta>0\,$ is a fixed number which may be chosen 
arbitrarily small.

In this domain we can integrate 
$\,\vert\partial_{p}^{\omega}\,G^{\,\alpha}_{l,s}\vert\,$ with
respect to $\,\vec \alpha\,$ and $\,\vec\lambda\,$ over the region indicated
in (\ref{int}) as long as we keep $\,\alpha\,$ finite. This could be shown 
by induction in $\,l+s\,$ using similar bounds as in (\ref{bon}):
\begin{displaymath}
\vert\partial_{p}^{\omega}\,G^{\,\alpha}_{l,s}(\vec\alpha,\vec\lambda,
p_{0,1}+ik_{0,1},\underline{p}_{1},\ldots,p_{0,n-1}+ik_{0,n-1},
\underline{p}_{n-1})\vert\
\end{displaymath}
\begin{equation}\label{bnd}
\leq 
\ e^{\,s\alpha 4m^{2}-m^{2}\sum_{j=1}^{s}\alpha_{j}}\,P
(\,\vert\vec p\vert\,)\,Q(\sqrt\alpha_{1},
\ldots,\sqrt\alpha_{s})\quad \mbox{for}
\quad \vec k_{0}\not=0\ .
\end{equation}       
For $\,\vec k_{0}=0\,$ we can use the bounds (\ref{bon}).

Therefore
\begin{equation}
\partial_{p}^{\omega}\,\Gamma_{l,s}^{\,\alpha,\,\alpha_{0}}(p_{0,1}+
ik_{0,1},\underline{p}_{1},\ldots,p_{0,n-1}+ik_{0,n-1},
\underline{p}_{n-1})
\end{equation}
is well-defined and the integrand is 
absolutely integrable.
Moreover we are able to give bounds for the  
$\,\vert\partial_{p}^{\omega}\,\Gamma_{l,s}^{\,\alpha,\,\alpha_{0}}\vert\,$ 
which for large $\,\alpha\,$ do not depend on $\,\alpha\,$ or $\,\alpha_{0}
\ $:
\vskip 0.3cm
{\bf Proposition 2.} {\it The integral representation of the
$\,\partial_{p}^{\omega}\,\Gamma_{l,s}^{\,\alpha,\,\alpha_{0}}\,$ in the 
domain defined by (\ref{img}) obeys the bounds}
\begin{displaymath}
\int\limits_{0}^{1}d\lambda_{1}
\ldots\int\limits_{0}^{1}d\lambda_{\sigma(l,s)}\int\limits_{
\alpha_{0}}^{\infty}d\alpha_{1}\ldots\int\limits_{\alpha_{0}}^{\infty}
d\alpha_{s}\,\vert\,\partial_{p}^{\omega}\,G_{l,s}^{\,\alpha}
(\vec \alpha,\vec
\lambda,p_{0,1}+ik_{0,1},\underline{p}_{1},\ldots,p_{0,n-1}+i
k_{0,n-1},\underline{p}_{n-1})\,\vert
\end{displaymath}
\begin{equation}\label{th1}
\leq
\cases{\ P_{1}(\,\vert\vec p\vert\,) & for 
$\ \alpha\geq\hat \alpha$ \cr
\ P_{2}(\,\vert\log(\alpha)
\vert\,)\ P_{3}(\,\vert\vec p\vert\sqrt\alpha\,)\ \alpha^{\,-2l+s+\frac{\vert
\omega\vert}{2}} & for 
$\ \alpha_{0}\leq \alpha \leq
\hat \alpha\quad.$\cr }
\end{equation}
\vskip 0.6 cm
{\it $ P_{k} \, $ are (each time they appear possibly new) polynomials with 
nonnegative coefficients which depend
neither on $\,\alpha\,$ nor on $\,\alpha_{0}\,$, but on $\,\eta,l,s,
\hat\alpha.\,$ $\,\hat \alpha>\alpha_{0}\,$ 
is some finite fixed number (e.g.1).} 
\vskip 0.3cm
{\it Proof.} The proof is performed by induction in $\,l+s\,$. We first
consider the case $\,\alpha\leq\hat\alpha\,$ and $\,s>2l\,$. Applying 
$\,\partial_{p}^{ 
\omega}\,$ on the recursive relation (\ref{gtl}), multiplying by $\, 
\Theta(\alpha-\alpha_{s})\,$
and taking the sum of the absolute values of all contributions leads to an 
inequality both sides of which can be integrated
with respect to $\,\vec\alpha\,$ and $\,\vec\lambda\,$ (due to the bounds 
(\ref{bnd})) over the domain 
indicated in (\ref{th1}). We change the order of integrations on the right
hand side and employ the induction hypothesis for 
$\,\alpha_{s}\leq\hat\alpha\,$. 
Furthermore we bound the powers of the $\,\alpha_{w_{k}}\,$ on the right 
hand side by powers of $\,\alpha_{s}\,$ and the corresponding $\,\alpha_{
w_{k}}$-integrals 
\begin{equation}\label{wkint}
\int\limits_{\alpha_{0}}^{\alpha_{s}}d\alpha_{w_{k}}\,e^{\,-\alpha_{w_{k}}
(\,q_{k}'^{2}+m^{2}-(\,\sum_{j=1}^{b_{1}+\ldots+b_{k}}k_{0,j})^{2}\,)}
\end{equation}
by $\,\alpha_{s}\cdot\,$constant uniformly in $\,0\leq\alpha_{0}<\alpha_{s}\leq\hat\alpha\,$.
We then obtain:
\begin{equation}
{\rm l.h.s.\  of\  (\ref{th1})}\ \leq\int\limits_{\alpha_{0}}^{\alpha}
d\alpha_{s}\int d^{\,4}p\,\Bigl\{\,e^{\,-\alpha_{s}p^{2}} 
\,\alpha_{s}^{\,-2l+s+1+\frac{\vert\omega\vert}{2}}\,P_{2}(\,\vert
\log(\alpha_{s})\vert\,)
\end{equation}
\begin{displaymath}
\qquad\mbox{    }\qquad P_{3}(\,\vert\vec p\vert\sqrt\alpha_{s},
\vert p_{0}\vert\sqrt\alpha_{s},\ldots,\vert p_{3}\vert\sqrt\alpha_{s}
\,)\,\Bigr \}\quad,
\end{displaymath}
where we have bounded appearing factors of the type $\,\vert k_{0,j}\vert
\sqrt\alpha_{s}\,$ by $\,2(m-\eta)\sqrt{\hat\alpha}\,$.
We now perform the loop-integration and get:
\begin{equation}
{\rm l.h.s.\  of\  (\ref{th1})}\ \leq\int\limits_{\alpha_{0}}^{\alpha}
d\alpha_{s}\,\alpha_{s}^{\,-2l+s-1+\frac{\vert\omega\vert}{2}}\,P_{2}(\,\vert
\log(\alpha_{s})\vert\,)\,P_{3}(\,\vert\vec p\vert\sqrt\alpha_{s}\,)\quad.
\end{equation}
Estimating the integral on the right hand side yields the induction 
hypothesis and completes the induction step.

The induction step for $\,s=2l\,$, $\,s=2l-1\,$ and 
$\,\alpha\leq\hat\alpha\,$ is also performed using (\ref{gtl}), and therefore
the argumentation is almost the same, but we have to split the contributions 
to (\ref{s2l}), (\ref{s1l}) with $\,\alpha_{s}\geq\alpha\,$ into two parts:
The first part with $\,\alpha_{s}\leq\hat\alpha\,$ can be treated as above.
For the second part we employ the induction hypothesis for $\,\alpha_{s}
>\hat\alpha\,$ on the right hand side of the inequality that we have 
obtained from (\ref{gtl}). 
(Note that $\,\vec p,\vec k_{0}=0\,$ for these terms.) 
We bound the $\,\alpha_{w_{k}}$-integrals (\ref{wkint}) by 
\begin{equation}
\int\limits_{0}^{\infty}d\alpha_{w_{k}}\,
e^{\,-\alpha_{w_{k}}m^{2}}\quad,
\end{equation}
compute the loop-integral, estimate the $\alpha_{s}$-integral and end up
with a constant independent of $\,\alpha,\alpha_{0}\,$. 
Taking all contributions together we thus can reproduce the induction
hypothesis.
This completes the induction step for the case $\,\alpha\leq
\hat\alpha\,$ which we can call the renormalizability part of Proposition 2. 

For the case $\,\alpha>\hat\alpha\,$ we first consider the induction step 
for $\,s>2l\,$. Looking at (\ref{gtl}) we examine the two types of 
contributions on the right hand side. The first type which has no reducible
line is easy to handle since the loop-integration variables can be left real
and therefore -- after applying $\,\partial_{p}^{\omega}\,$, taking the 
absolute value, multiplying by $\,\Theta(\alpha-\alpha_{s})\,$,  
integrating over the region indicated in (\ref{th1}) and employing the 
induction hypothesis for $\,\alpha_{s}\leq\hat\alpha\,$ and for 
$\,\alpha\geq\alpha_{s}>\hat\alpha\,$ -- we can reproduce the induction 
hypothesis for $\,\alpha>\hat\alpha\,$ without any difficulties. 
The second type which can be described
graphically as a sum of chains that consist of one-particle irreducible
Feynman graphs which are connected to their neighbours by a reducible line
requires a more careful analysis, because the propagators corresponding to
the reducible lines may increase exponentially in the $\,\alpha_{w_{k}}\,$.
Due to this fact we will have to add an imaginary part to the 
loop-integration variable $\,p_{0}\,$ that means instead of integrating 
along the real $\,p_{0}$-axis we want to integrate along the path $\,
p_{0}+ik_{0}\,$ with fixed $\,k_{0}\,$. (We are free to do so because 
(\ref{kog}) is polynomially bounded.)
Let us now have a look at the exponent
of a propagator corresponding to a reducible line. In order to get an 
exponential decrease we have to achieve
\begin{equation}\label{exp}
\Bigl [\,m^{2}+(\,\underline{p}+\sum_{j=1}^{b_{1}+\ldots+b_{k}}
\underline{p}_{j})^{2}+(\,p_{0}+\sum_{j=1}^{b_{1}+\ldots+b_{k}}p_{0,j})^{2}
-(k_{0}+\sum_{j=1}^{b_{1}+\ldots+b_{k}}k_{0,j})^{2}\Bigr ]\, \geq\, 
\varepsilon\, >\, 0\ .
\end{equation}
Furthermore the real part of the exponent of the differentiated 
propagator which corresponds to the line that closes the loop has to be 
negative to get an exponential decrease in $\,\alpha_{s}\,$:
\begin{equation}\label{lex}
(\,m^{2}+\underline{p}^{2}+p_{0}^{2}-k_{0}^{2})\,\geq\,\varepsilon\,>\,0
\end{equation}
and therefore we require
\begin{equation}\label{k01}
\vert k_{0}\vert\ \leq\ m-\eta\quad.
\end{equation}
We now want to show that for each fixed chain that
means for each contribution to the second term of (\ref{gtl}) with fixed 
$\,v,c_{1},\ldots,c_{v},d_{1},\ldots,d_{v}\,$ and a fixed order of 
external momenta $\,p_{1},\ldots,p_{n-1}\,$  
we can find an imaginary part $\,k_{0}\,$ for the loop-integration that 
fulfills (\ref{k01}) and (\ref{exp}) for all $\,k=1,\ldots,v-1\,$.
For a fixed chain we define
\begin{equation}\label{avi}
\hat q_{k}:=\sum_{j=1}^{b_{1}+\ldots+b_{k}}k_{0,j}\quad,\quad k=1,\ldots,v-1
\quad. 
\end{equation}
(\ref{img}) implies
\begin{equation}\label{bav}
\vert\hat q_{k}\vert \leq 2(m-\eta)\quad\mbox{ and }\quad
\vert\hat q_{k}-\hat q_{i}\vert\leq 2(m-\eta)\quad\mbox{ for all}\quad k,i
\quad.
\end{equation}
Now it is easy to realize
\vskip 0.3cm
{\bf Lemma 3.} {\it Let $\,k_{0}\,$ be a real number bounded by}
\begin{equation}\label{k02}
-\min\{0,\min_{k}\{\hat q_{k}\}\}-m+\eta\leq k_{0}\leq-\max\{0,\max_{k}
\{\hat q_{k}\}\}+m-\eta\quad,
\end{equation}
{\it then $\,k_{0}\,$ fulfills (\ref{k01}) and (\ref{exp}) for all k.}
\vskip 0.3cm
{\it Proof.} Because of (\ref{bav}) we can always find a $\,k_{0}\,$ which 
obeys 
(\ref{k02}), and from (\ref{k02}) we get 
\begin{equation}\label{ak0}
\vert\hat q_{k}+k_{0}\vert\ \leq\ m-\eta\quad\mbox{ for all }\ k\quad,
\end{equation}
and therefore we can see that for this
$\,k_{0}\,$ (\ref{k01}) and (\ref{exp}) are fulfilled for all $\,k\,$.
$\,\Box\,$

Since we want to employ the induction hypothesis on the right hand side of
(\ref{gtl}) $\,k_{0}\,$ has to satisfy another condition: 

$\,(k_{0},k_{0,1},\ldots,k_{0,b_{1}}),(k_{0}+\hat q_{1},
k_{0,i_{2}+1},\ldots,k_{0,i_{2}+b_{2}}),\ldots,(k_{0}+\hat q_{v-1},
k_{0,i_{v}+1},\ldots,k_{0,n-1},-k_{0})\,$ \\$\,(i_{k}=\sum_{j=1}^{k-1}b_{j})
\,$ 
have to be in the domain indicated by (\ref{img}). Therefore we have to
modify the bounds which $\,k_{0}\,$ has to obey. We define
\begin{equation}\label{mht}
\hat m_{k}:=\max_{\tau_{a,k}}\{\hat q_{k-1}+\sum_{j\in\tau_{a,k}}
k_{0,i_{k}+j}\}\quad,\quad \check m_{k}:=\min_{\tau_{a,k}}\{\hat q_{k-1}+
\sum_{j\in\tau_{a,k}}k_{0,i_{k}+j}\}\quad,\quad k=1,\ldots,v\quad,
\end{equation}
where
\begin{displaymath}
\hat q_{0}:=0\quad,\quad \tau_{a,k}\subseteq\{1,\ldots,b_{k}\}\quad.
\end{displaymath}

Furthermore we define
\begin{equation}\label{max}
\hat m:=\max_{k}\{\hat m_{k}\}\quad\mbox{ and }\quad \check m:=\min_{k}\{
\check m_{k}\}\quad.
\end{equation}
(\ref{img}) implies
\begin{equation}
\hat m\ \leq\ 2(m-\eta)\quad\mbox{ and }\quad \check m\ \geq\ -2(m-\eta)
\quad. 
\end{equation}
\newpage
Now we are ready to prove

{\bf Lemma 4.} {\it Let $\,k_{0}\,$ be a real number bounded by}
\begin{equation}\label{k03}
k_{0}\ \leq\ \min\cases{\ 2(m-\eta)-\max\{0,\hat m\} & \cr
\ m-\eta-\max\{0,\max_{\,k\,}\{\hat q_{k}\}\} & \cr}
\end{equation}
\begin{displaymath}
k_{0}\ \geq\ \max\cases{\ -2(m-\eta)-\min\{0,\check m\} & \cr
\ -m+\eta-\min\{0,\min_{\,k\,}\{\hat q_{k}\}\} & \cr}\quad,
\end{displaymath}
{\it then $\,k_{0}\,$ fulfills (\ref{k01}), (\ref{exp}) and}
\begin{equation}\label{k04}
\vert k_{0}+\hat q_{k-1}+\sum_{j\in\tau_{a,k}}k_{0,i_{k}+j}\vert\ \leq\ 2(m-
\eta)\quad
\mbox{ {\it for all} }\quad \tau_{a,k}\subseteq\{1,\ldots,b_{k}\}\quad,\quad
k=1,\ldots,v\quad.
\end{equation}
{\it Proof.} Due to (\ref{img}) we can always find a $\,k_{0}\,$ that obeys 
(\ref{k03}).
It is not difficult to check that this $\,k_{0}\,$ -- besides fulfilling
(\ref{k01}), (\ref{exp}) due to Lemma 3 -- also fulfills (\ref{k04}). $\,
\Box\,$

Now we are able to carry out the induction step for $\,s>2l\,$:
For every chain on the right hand side of (\ref{gtl}) we choose a 
corresponding $\,k_{0}\,$ which satisfies (\ref{k01}), (\ref{exp}) for all
$\,k\,$ and (\ref{k04}). Then we apply $\,\partial_{p}^{\omega}\,$, take 
the absolute value, multiply by $\,\Theta(\alpha-\alpha_{s})\,$, 
integrate over the domain indicated in (\ref{th1}) and
employ the induction hypothesis for $\,\alpha_{s}\leq\hat\alpha\,$ -- for
this contribution we can refer to the case $\,\alpha\leq\hat\alpha\,$ treated
above -- and
for $\,\alpha\geq\alpha_{s}>\hat\alpha\,$ on the right hand side. We
bound the $\,\alpha_{w_{k}}$-integrals 
\begin{displaymath}
\int\limits_{\alpha_{0}}^{\alpha_{s}}d\alpha_{w_{k}}\,e^{-\alpha_{w_{k}}
(\,q_{k}'^{2}+m^{2}-(k_{0}+\hat q_{k})^{2})}\ldots
\end{displaymath}
by 
\begin{displaymath}
\int\limits_{0}^{\infty}d\alpha_{w_{k}}\,e^{\,-\alpha_{w_{k}}
\eta(2m-\eta)}\ldots\quad 
\end{displaymath}
and end up with
\begin{equation}
{\rm l.h.s\ of \ (\ref{th1})}\ \leq\ P_{1}(\,\vert\vec p\vert\,)
+\int\limits_{\hat\alpha}^{\alpha}d\alpha_{s}\Bigl \{\int d^{\,4}p
\ e^{\,-\alpha_{s}(\,p^{2}+m^{2})}\,P_{1}(\,\vert\vec p\vert,\vert p_{0}\vert,
\ldots,\vert p_{3}\vert\,)
\end{equation}
\begin{displaymath}
+\sum\int d^{\,4}p\ e^{\,-\alpha_{s}(\,p^{2}-k_{0}^{2}+m^{2})}\,P_{1}(\,\vert
\vec p\vert,\vert p_{0}\vert,\ldots,\vert p_{3}\vert\,)\,\Bigr \}\quad.
\end{displaymath}
Now we perform the loop-integration, set $\,\alpha=\infty\,$ and since
$\,m^{2}-k_{0}^{2}\geq\eta(2m-\eta)>0\,$ we obtain the induction hypothesis 
for
$\,\alpha>\hat \alpha\,$.

For (\ref{s2l}) and (\ref{s1l}) the argumentation is almost the same.
Note in particular that $\,\lambda_{j}\in[0,1]\,$ for all $\,j=1,\ldots,
\sigma(l,s)\,$ so that if (\ref{img}) is fulfilled for $\,\vec k_{0}\,$ it 
also holds for $\,\lambda_{j}\vec k_{0}\,$.
This completes the proof of Proposition 2. $\,\Box\,$

Because $\,\alpha\,$ and $\,\alpha_{0}\,$ only appear as bounds of the domain
of integration in (\ref{int}) and (\ref{th1}), we also get from Proposition 2
the convergence of the integral representation of $\,\partial_{p}^{\omega}\,
\Gamma^{\,\alpha,\,\alpha_{0}}_{l,s}\,$ as $\,\alpha_{0}\rightarrow 0\,$ and
$\,\alpha\rightarrow\infty\,$. Thus we obtain
\newpage
{\bf Theorem 5.} {\it The one-particle irreducible renormalized Green 
functions of perturbative Euclidean massive $\,\Phi_{\,4}^{\,4}\,$}
\begin{equation}
\Gamma_{l,s}^{\,\infty,\,0}(p_{0,1}+
ik_{0,1},\underline{p}_{1},\ldots,p_{0,n-1}+ik_{0,n-1},
\underline{p}_{n-1})
\end{equation}

{\it are analytic in $\,p_{0,1}+ik_{0,1},\ldots,p_{0,n-1}+ik_{0,n-1}\,$ 
in the domain defined by}

\begin{displaymath}
(p_{0,1},\ldots,p_{0,n-1})\in {\rm\bf R}^{n-1}\quad\mbox{ {\it and} }
\quad(k_{0,1},\ldots,k_{0,n-1})\in {\rm\bf R}^{n-1} 
\end{displaymath}
\begin{equation}\label{th5}
\mbox{{\it with} }\qquad 
\vert\,\sum_{j\in\tau_{a}}k_{0,j}\,\vert<2m\quad\mbox{  {\it for all}  }\quad
\tau_{a}\subseteq\{1,\ldots,n\}\quad,\quad k_{0,n}=-\sum_{j=1}^{n-1}k_{0,j}
\end{equation}

{\it and smooth with respect to $\,(\underline{p}_{1},\ldots,
\underline{p}_{n-1})\in{\rm\bf R}^{3(n-1)}\,$.
The integrands of their integral representations (\ref{int}) are absolutely
integrable.}
\section{Structure of the Integrands $\,G_{l,s}^{\,\alpha}\,$}
Using (\ref{sgl}), (\ref{gtl}), (\ref{s2l}) and (\ref{s1l}) we now want 
to analyse the structure of the integrands $\,G_{l,s}^{\,\alpha}\,$. 
We can state  
\vskip 0.3cm
{\bf Proposition 6.} {\it The integrands $\,G_{l,s}^{\,\alpha}\,$ of the 
renormalized one-particle irreducible Green functions 
$\,\Gamma_{l,s}^{\,\alpha,\,0}\,$ with the renormalization
conditions (\ref{ren}) have the following structure:}

\begin{equation}\label{ga0}
G^{\,\alpha}_{l,s}(\vec \alpha,\vec \lambda,\vec
p\,)\ =\ \sum_{j}\,P_{j}(\,\vec p\,)
\,Q_{j}
(\vec \alpha,\vec
\lambda)\,e^{\,-\sum_{k,v=1}^{n-1}\,A^{\,j}_{kv}(\vec\alpha,\vec
\lambda)\, p_{k}p_{v}}
\,V_{j}^{\,\alpha}(\vec \alpha)
\,e^{\,-m^{2}\sum_{k=1}^{s}\alpha_{k}}\quad.
\end{equation}

{\it{\rm (a)} $\,V_{j}^{\,\alpha}(\vec \alpha)\,$ are products of $\,
\Theta$-functions
in $\,(\alpha_{i}-\alpha_{k})\ ,\ \pm(\alpha-\alpha_{s})\,$. The support of
$\,V_{j}^{\,\alpha}(\vec\alpha)\,$ restricts all $\,\alpha_{i}\,$ appearing
as arguments of $\,A_{kv}^{\,j}(\vec\alpha,\vec\lambda)\,$ to
$\,\alpha_{i}\leq\alpha\,$.

{\rm (b)} $\,A^{\,j}_{kv}(\vec \alpha,\vec \lambda)\, $ are continuous 
with respect to $\,\vec \lambda\,$ and with respect
to $\,\vec \alpha\, $ in the support of $\,V_{j}^{\,\alpha}
(\vec \alpha)\,$. They are homogeneous of degree 1 in $\,\vec\alpha\,$ 
(that means $\,A^{\,j}_{kv}(\tau\vec\alpha,\vec\lambda)=\tau\,
A^{\,j}_{kv}(\vec\alpha,\vec\lambda)\,$),
and $\,A^{\,j}(\vec\alpha,\vec\lambda)\,$ is a positive semi-definite 
symmetrical $\,n-1\times n-1\,$ matrix.

{\rm (c)} $\,Q_{j}(\vec \alpha,\vec \lambda)\,$ are rational functions in
$\,\vec \alpha\,$ and $\,\vec\lambda\,$ which are homogeneous of degree
$\,d_{j}\in{\rm\bf Z}\,$ in $\,\vec\alpha\,$.
 
{\rm (d)} $\,P_{j}(\vec p)=\prod_{\,k\leq v}\,(p_{k}p_{v})^{\,
u_{k,v}^{\,j}}\ ,\ u_{k,v}^{\,j}\in {\rm\bf N}_{0}\,$, are monomials
in $\,O(4)$-invariant scalar products of the $\,p_{i}\,$.

All functions introduced on the right hand side of (\ref{ga0}) also depend
on $\,l,s\,$ and the sum over $\,j\,$ is finite.}
\vskip 0.5cm 
{\it Proof.} The proof is carried out by induction in $\,l+s\,$.
We employ the induction hypothesis on the right hand side of (\ref{gtl}).
First looking at the second term on the right hand side we realize 
that this contribution can again be written as a sum
of terms of the form (\ref{ga0}). For these terms and also for the first
term on the right hand side of (\ref{gtl})
we thus
obtain loop-integrals of the type
\begin{equation}
\int d^{\,4}p\ P_{j}(\,\vec p\,)\ e^{\,-\alpha_{s}\,p^{2}\,
-\sum_{k,v=1}^{n+1}\,A^{\,j}_{kv}\,p_{k} p_{v}}\quad,
\end{equation}
where (according to (\ref{gtl})) $\ p_{n+1}=-p_{n}=-p\ $. 

Using
\begin{equation}\label{ajkv}
-\alpha_{s}\,p^{2}
-\sum_{k,v=1}^{n+1}\ A^{\,j}_{kv}\ p_{k}p_{v}
=-\sum_{k,v=1}^{n-1}\ \biggl [\ A_{kv}^{\,j}-
\frac{(\,A_{kn}^{\,j}-A_{kn+1}^{\,j}\,)\ (\,A_{vn}^{\,j}-A_{vn+1}^{\,j}\,)}
{A_{nn}^{\,j}
+A_{n+1n+1}^{\,j}-2A_{nn+1}^{\,j}+\alpha_{s}}\ \biggr ]\,p_{k}p_{v}
\end{equation}

\begin{displaymath}
-(\,A_{n+1n+1}^{\,j}+A_{nn}^{\,j}-2A_{nn+1}^{\,j}+\alpha_{s}\,)\biggl (\,p+
\sum_{k=1}^{n-1}\ \frac{A_{kn}^{\,j}-A_{kn+1}^{\,j}}{A_{nn}^{\,j}
+A_{n+1n+1}^{\,j}-2A_{nn+1}^{\,j}+\alpha_{s}}\ p_{k}\,\biggr )^{\,2}
\end{displaymath}

we can perform the loop-integration and reproduce the induction hypothesis
for $\,\tilde G_{l,s}\,$. Note that $\,\alpha_{s}\geq\alpha_{i}\,$ due to
the induction hypothesis, see in particular (a). Therefore the second part
in the square brackets in (\ref{ajkv}) is always well-defined. Furthermore
the new contributions to the $\,P_{j}(\vec p)\,$ are again of the form
indicated in (d) which could be seen by using
\begin{equation}
(p\,p)^{\,\tilde u}\,(p\,p_{k})^{\,\hat u}\,e^{\,p\,y}=(\nabla_{y}\,
\nabla_{y})^{\,
\tilde u}\,(\nabla_{y}\,p_{k})^{\,\hat u}\,e^{\,p\,y}
\end{equation}
before computing the Gaussion integral and setting $\,y=0\,$
afterwards.
Inserting $\,\tilde G_{l,s}\,$ in (\ref{sgl}), (\ref{s2l}) and (\ref{s1l})
again yields the induction hypothesis and completes the induction step.
Note that the $\,A^{\,j}$'s for 0-momentum $\,\tilde G$'s in (\ref{s2l}),
(\ref{s1l}) are simply defined to be 0. $\,\Box\,$

In order to bound the degree of homogeneity $\,d_{j}\,$ of 
$\,Q_{j}(\vec \alpha,\vec\lambda)\,$ we define for fixed $\,l,s\,$

\begin{equation}
h_{j}\ :=\ 2\,\sum_{k\leq v}\,u_{k,v}^{\,j}\quad.
\end{equation}
Inserting (\ref{ga0}) in (\ref{sgl}), (\ref{gtl}), (\ref{s2l}) and (\ref{s1l})
we can prove by induction in $\,l+s\,$ that the following equation
holds:
\begin{equation}
\frac{1}{2}\,h_{j}\,-\,d_{j}\ =\ 2\,l\quad,\quad\mbox{   for all }\ l,s,j
\quad.
\end{equation}
We thus obtain for
$\,G_{l,s}^{\,\infty}\,$ and for all $\,l,s,j\,$
\begin{equation}\label{djs}
d_{j}\ >\ -\,s \quad.
\end{equation}
For $\,s>2l\,$ this is obvious, and for $\,s=2l\,$ we see in (\ref{s2l})
that because $\,\alpha=\infty\,$ the first contribution is vanishing, and 
with the help of (\ref{ga0}) we conclude that therefore only terms with 
$\,h_{j}\geq 2\,$ contribute. For $\,s=2l-1\,$ we get from (\ref{s1l}) 
and (\ref{ga0}) that for $\,\alpha=\infty\,$
only terms with $\,h_{j}\geq 4\,$ contribute.

Now it is easy to prove
\vskip 0.3cm
{\bf Corollary 7.} {\it The integral representation (\ref{int}) of the 
one-particle irreducible renormalized Green functions $\,\Gamma_{l,s}^{\,
\infty,\,0}\,$ with the renormalization conditions (\ref{ren}) can be 
written as}
\begin{displaymath}
\Gamma^{\,\infty,\,0}_{l,s}(\,\vec
p\,)\,=\int\limits_{0}^{1}
d\vec\lambda\int\limits_{0}^{1}d\beta_{1}\ldots\int\limits_{0}^{1}
d\beta_{s}\,\delta\Bigl(1-\sum_{k=1}^{s}\beta_{k}\Bigr)\Biggl [\,\sum_{j}
\,V_{j}(\vec\beta\,)\,P_{j}(\,\vec p\,)\,Q_{j}(\vec\beta,\vec\lambda)
\end{displaymath}
\begin{equation}\label{pr7}
\frac{1}
{\Bigl (\,\sum_{k,v}A^{j}_{kv}(\vec\beta,\vec
\lambda)\,p_{k}p_{v}+\,m^{2}\sum_{k=1}^{s}\beta_{k}\Bigr )^{\,d_{j}+s}}
\ \Biggr ]\quad,
\end{equation}
\vskip 0.3cm
{\it and the integrand is absolutely integrable. The momentum derivatives 
$\,\partial_{p}^{\omega}\,
\Gamma_{l,s}^{\,\infty,\,0}\,$ are represented by integrals of the 
corresponding momentum derivatives of the integrand which are also absolutely
integrable, and due to Theorem
5 this representation is still valid in the complex domain
indicated in (\ref{th5}).}
\vskip 0.5cm
{\it Proof.} Considering the integral representation (\ref{int}) of the
renormalized one-particle irreducible Green functions we define a 
substitution of the integration
variables (we are free to do so because the integrand is absolutely
integrable) by
\begin{equation}
\alpha_{k}=:\tau\beta_{k}\quad,\quad k=1,\ldots,s\quad\mbox{ and
}\quad\sum_{k=1}^{s}\beta_{k}=1\quad.
\end{equation} 
This yields
\begin{equation}
d\alpha_{1}\ldots d\alpha_{s}\ =\ \tau^{s-1}\,\delta\Bigl(\,1-\sum_{k=1}^{s}
\beta_{k}\,\Bigr)\,d\beta_{1}\ldots d\beta_{s}d\tau\quad.
\end{equation}

Now we insert the integrand from Proposition 6 and get

\begin{displaymath}                  
\Gamma^{\,\infty,\,0}_{l,s}(\,\vec p\,)\,=\int\limits_{0}^{1}
d\vec\lambda\int\limits_{0}^{1}d\beta_{1}
\ldots\int\limits_{0}^{1}d\beta_{s}\int\limits_{0}^{\infty}d\tau
\,\biggl [ \delta\Bigl(1-\sum_{k=1}^{s}\beta_{k}\Bigr)\,\sum_{j}\,V_{j}
(\vec\beta\,)\,P_{j}(\,\vec p\,)
\end{displaymath}
\begin{equation}
\tau^{s+d_{j}-1}
\,Q_{j}(\vec\beta,\vec \lambda)\,e^{\,-\tau\sum_{k,v}A^{j}_{kv}
(\vec\beta,\vec\lambda)\,p_{k}p_{v}}\,e^{\,-\tau\,m^{2}\sum_{k=1}^{s}
\beta_{k}}\biggr ]\quad.
\end{equation}

Due to (\ref{djs}) and because the integrand is absolutely
integrable we can perform the \\
$\,\tau$-integration and then we obtain (\ref{pr7}). $\,\Box\,$

\section{Relativistic massive $\,\Phi_{\,4}^{\,4}\,$}
\subsection{The regularized Theory}
We now want to turn our attention to the corresponding relativistic theory.
We define a relativistic regularized propagator which is analytic in momentum
space by
\begin{equation}\label{rpr}
\tilde C_{\alpha}^{\,\alpha_{0}}(p):=\int\limits_{\alpha_{0}}^{\alpha}
d\alpha'\,e^{\,-\alpha'(\,p\eta p+(\varepsilon+i)m^{2})}\quad,\qquad
\varepsilon>0\ ,\ 0<\alpha_{0}\leq\alpha<\infty\quad,
\end{equation}
where $\,\eta\,$ is the matrix
\begin{equation}\label{eta}
\eta:=\left( \begin{array}{cccc}
\varepsilon-i & 0 & 0 & 0 \\
0 & \varepsilon+i & 0 & 0 \\
0 & 0 & \varepsilon+i & 0 \\ 
0 & 0 & 0 & \varepsilon+i
\end{array}\right )\qquad.
\end{equation}
The interaction Lagrangian at scale $\,\alpha_{0}\,$ is defined by
\begin{displaymath}
L^{\,\alpha_{0},\,\alpha_{0}}(\Phi):=\sum_{r\geq 1}g^{\,r}\,
L_{r}^{\,\alpha_{0},\,\alpha_{0}}(\Phi)
\end{displaymath}
and
\begin{equation}\label{lre}
L_{r}^{\,\alpha_{0},\,\alpha_{0}}(\Phi):=
\int d^{\,4}x\ \left(\,a_{r}^{\,\alpha_{0}}\Phi^{\,2}(x)+b_{r}^{\,\alpha_{0}}
 \Phi(x)\Delta\Phi(x)-d_{r}^{\,\alpha_{0}}\Phi(x)
\partial_{0}^{2}\Phi(x)+c_{r}^{\,\alpha_{0}}\Phi^{\,4}(x)\right)\quad,
\end{equation}
$\,\Delta\,$ denotes the 3-dim Laplace operator. As the $\,
\varepsilon$-regularization breaks Lorentz invariance (but not $\,O(3)$-
and T-invariance) the interaction
Lagrangian contains an additional counterterm.

In analogy to the Euclidean theory the effective Lagrangian
\begin{displaymath}
L^{\,\alpha,\,\alpha_{0}}(\Phi):=\sum_{r\geq 1}g^{\,r}\,
L_{r}^{\,\alpha,\,\alpha_{0}}(\Phi)
\end{displaymath}
is defined through
\begin{equation}\label{rla}
e^{\
i\,L^{\,\alpha,\,\alpha_{0}}(\Phi)\ +i\,I^{\,\alpha,\,\alpha_{0}}}:=\ 
e^{\,\Delta(\alpha,\alpha_{0})}\
e^{\,i\,L^{\alpha_{0},\,\alpha_{0}}(\Phi)}\quad ,
\end{equation}
where the functional Laplace operator $\,\Delta(\alpha,\alpha_{0})\,$ is
defined as in section 2 but with the relativistic propagator (\ref{rpr}).

Differentiating (\ref{rla}) with respect to $\,\alpha\,$ yields the
flow equation for the effective Lagrangian of the relativistic
theory:
\begin{equation}\label{rfl}
\partial_{\alpha}\,L^{\,\alpha,\,\alpha_{0}}(\Phi)\, +\, 
\partial_{\alpha}\,I^{\,\alpha,\,\alpha_{0}}\ =\ 
\left[\,\partial_{\alpha}\Delta(\alpha,\alpha_{0})\,\right]\,
L^{\,\alpha,\,\alpha_{0}}(\Phi)
\end{equation}

\begin{displaymath}
+\frac{i}{2}\int\,d^{\,4}x\int\,d^{\,4}y
\,\left (\frac{\delta}{\delta\,\Phi(x)}\,L^{\,\alpha,\,\alpha_{0}}(\Phi
)\right)\left(\partial_{\alpha}\,C_{\alpha}^{\,\alpha_{0}}(x-y)\right )
\frac{\delta}{\delta\,\Phi(y)}\,L^{\,\alpha,\,\alpha_{0}}(\Phi)\quad.
\end{displaymath}
\subsection{Flow Equations for one-particle irreducible Green functions}
The generating functional $\,W_{c}^{\,\alpha,\,\alpha_{0}}(J)\,$ of the 
pertubative, regularized connected Green functions of the
relativistic theory is given by
\begin{equation}\label{wcr}
W_{c}^{\,\alpha,\,\alpha_{0}}(J):=i\,L^{\,\alpha,\,\alpha_{0}}(\Phi)\,\vert
_{\,\Phi=i\tilde C_{\alpha}^{\,\alpha_{0}}\!J}+i\,I^{\,\alpha,\,
\alpha_{0}}-\frac{1}{2}<J,\tilde C_{\alpha}^{\,\alpha_{0}}J>\quad .
\end{equation}
Then the generating functional $\,i\,\Gamma^{\,\alpha,\,\alpha_{0}}(\Phi_{c})
\,$ of the corresponding one-particle irreducible Green functions is defined 
by
\begin{equation}\label{gar}
i\, \Gamma^{\,\alpha,\,\alpha_{0}}(\Phi_{c}):=
\Bigl [ \,W_{c}^{\,\alpha,\,\alpha_{0}}(J) -i < J ,\Phi_{c}>\,   
\Bigr ]_{\,J=J(\Phi_{c})} \quad,
\end{equation}
where
\begin{equation}\label{phr}
\Phi_{c}(p,J)=\frac{1}{i}(2\pi)^{\,4}\delta_{J(-p)}\,W_{c}^{\,\alpha,\,
\alpha_{0}}
(J)=\tilde C_{\alpha}^{\,\alpha_{0}}(p)\,\Bigl \{\,(2\pi)^{\,4}
\,\delta_{\Phi
(-p)}\,i\,L^{\,\alpha,\,\alpha_{0}}(\Phi)\,\vert_{\,\Phi=i\tilde C_{\alpha}
^{\,\alpha_{0}}\!J}\,+i\,J(p)\,\Bigr \}\ .
\end{equation}
Differentiating (\ref{gar}) with respect to $\,\alpha\,$ and using 
(\ref{wcr}), (\ref{rfl}) and (\ref{phr}) 
we get
\begin{equation}\label{fr1}
\partial_{\alpha}\,(\,i\,\Gamma^{\,\alpha,\,\alpha_{0}}(\Phi_{c})+
\frac{1}{2}<\Phi_{c},\{\tilde C_{\alpha}^{\,\alpha_{0}}\}^{\,-1}\Phi_{c}>
 )=[\partial_{\alpha}\,\tilde\Delta(\alpha,\alpha_{0})\,]
\ i\,L^{\,\alpha,\,\alpha_{0}}(\Phi)\,\vert_{\,\Phi=i\tilde C_{\alpha}
^{\,\alpha_{0}}\!J(\Phi_{c})}\quad .
\end{equation}
We define
\begin{equation}\label{hgr}
\hat\Gamma_{r}^{\,\alpha,\,\alpha_{0}}(q,p,\Phi_{c})\ :=\ 
(2\pi)^{\,4}\left \{\,\delta_{\Phi(p)}\delta_{\Phi(q)}\,L^{\,\alpha,\,
\alpha_{0}}(\Phi)\,\vert_{\,\Phi=i\tilde C_{\alpha}^{\,\alpha_{0}}\!J(
\Phi_{c})}\right \}_{r}\quad .
\end{equation}
By the same procedure as in section 3.2
((\ref{help1}),$\ldots$,(\ref{re2})) we now  
obtain a recursive
relation 

for $\,\hat\Gamma_{r}^{\,\alpha,\,\alpha_
{0}}(q,p,\Phi_{c})\,$ that allows us to express $\hat\Gamma_{r}^{\,
\alpha,\,\alpha_{0}}(q,p,\Phi_{c})$ in terms of $\Gamma_{k}^{\,\alpha,
\,\alpha_{0}}(\Phi_{c})\,,\,k=1,\ldots,r$

\begin{equation}\label{rer}
\hat \Gamma_{r}^{\,\alpha,\,\alpha_{0}}(q,p,\Phi_{c})=(2\pi)^{\,4}\,
\delta_{\Phi_{c}(p)}\delta_{\Phi_{c}(q)}\,\Gamma_{r}^{\,\alpha,\,\alpha_{0}}
(\Phi_{c})
\end{equation}
\begin{displaymath}
+\,i(2\pi)^{\,4}\sum_{k=1}^{r-1}\int d^{\,4}q'\ \tilde C_{\alpha}^{\,
\alpha_{0}}(q')\ \hat\Gamma_{r-k}^{\,
\alpha,\,\alpha_{0}}(q,-q',\Phi_{c})\ \delta_{\Phi_{c}(p)}\delta_{\Phi_{c}
(q')}\,\Gamma_{k}^{\,\alpha,\,\alpha_{0}}(\Phi_{c})\quad .
\end{displaymath}
Using (\ref{fr1}), (\ref{hgr}) and (\ref{rer}) the differential flow
equation for the relativistic one-particle irreducible Green
functions
$\,i\,\Gamma_{r,n}^{\,\alpha,\,\alpha_{0}}(p_{1},\ldots,p_{n-1})\,$
reads:
\begin{equation}\label{fr2}
\partial_{\alpha}\,i\,\Gamma_{r,n}^{\,\alpha,\,\alpha_{0}}(p_{1},\ldots,
p_{n-1})=\frac{1}{2}\int\frac{d^{\,4}p}{(2\pi)^{\,4}}\ (\partial_{\alpha}\,
\tilde
C_{\alpha}^{\,\alpha_{0}}(p))\ i\,\hat \Gamma_{r,n+2}^{\,\alpha,\,
\alpha_{0}}(p,-p,p_{1},\ldots,p_{n-1})\quad,
\end{equation}
where
\begin{equation}\label{ker}
i\,\hat\Gamma_{r,n+2}^{\,\alpha,\,\alpha_{0}}(p,-p,p_{1},\ldots,p_{n-1}):=
(n+1)(n+2)\,i\,\Gamma_{r,n+2}^{\,\alpha,\,\alpha_{0}}(p,-p,p_{1},\ldots,
p_{n-1})
\end{equation}
\begin{displaymath}
+\sum_{v=2}^{r}\sum_{\{a_{j}\},\{b_{j}\}}
K^{\,v}(b_{1},\ldots,b_{v})\Bigl [\,\prod_{k=1}^{v-1}\,\tilde C_{\alpha}^{\,
\alpha_{0}}(q_{k}')\ i\,\Gamma_{a_{k},b_{k}+2}^{\,\alpha,\,\alpha_{0}}
(q_{k-1}',p_{i_{k}+1},\ldots,p_{i_{k}+b_{k}})
\end{displaymath}
\begin{displaymath}
i\,\Gamma_{a_{v},b_{v}+2}^{\,\alpha,\,\alpha_{0}}(q_{v-1}',-p,p_{i_{v}+1},
\ldots,p_{n-1})\,\Bigr ]_{\,{\rm symm.}}\quad.
\end{displaymath}
The notation is the same as before (see (\ref{ket})). 

Looking at (\ref{re2}), (\ref{fg4}) and (\ref{ket}) we realize that 
multiplying
the Euclidean one-particle irreducible Green functions by $\,(-1)\,$ yields a 
flow equation which is -- apart from the different definition of the 
propagators -- identical 
to the flow equation for the relativistic one-particle irreducible Green 
functions. 

Regarding the propagators we observe that we have to
replace the Euclidean metric by the matrix $\,\eta\,$ and that we
have to add a factor $\,(\varepsilon+i)\,$ to the mass term
$\,m^{2}\,$ in order to get the relativistic propagator. Therefore it
turns out that as long as we keep $\,\varepsilon > 0\,$ we can easily
transfer most of our results from the Euclidean theory to the 
new situation.      

\subsection{Integral Representation and Renormalizability}
The boundary values at $\,\alpha=\alpha_{0}\,$ follow from
(\ref{lre}) and read:
\begin{equation}\label{a0a}
\Gamma_{r,2}^{\,\alpha_{0},\,\alpha_{0}}(p)\ =\ a_{r}^{\,\alpha_{0}}-
b_{r}^{\,\alpha_{0}}\underline{p}^{2}+d_{r}^{\,\alpha_{0}}p_{0}^{2}\quad ,
\quad 
\Gamma_{r,4}^{\,\alpha_{0},\,\alpha_{0}}(p_{1},p_{2},p_{3})=c_{r}^{\,
\alpha_{0}}\quad,\quad \Gamma_{r,n}^{\,\alpha_{0},\,\alpha_{0}}(\vec p)
\equiv 0\quad\mbox{for}\ n>4\ .
\end{equation}

This implies (as in the Euclidean theory)
\begin{equation}\label{war}
\partial_{p}^{\omega}\,\Gamma_{r,n}^{\,\alpha_{0},\,\alpha_{0}}(\vec
p)\equiv 0\quad\mbox{for}\quad n+\vert\omega\vert>4\quad.
\end{equation}

Changing the indices from $\,(r,n)\,$ to $\,(l,s)\,$ (see (\ref{ls})) and 
using renormalization conditions which correspond to (\ref{ren})
multiplied by $\,(-1)\,$
\begin{equation}\label{renr}
i\,\Gamma_{0,0}^{\,\infty,\,\alpha_{0}}(0):=-c_{1}^{\,R}\quad\mbox{and}
\quad
\Gamma_{l,2l}^{\,\infty,\,\alpha_{0}}(0)=0\,,\,\Gamma_{l,2l-1}^{\,\infty,\,
\alpha_{0}}(0)=0\,,\,\partial_{\mu}\,\partial_{\nu}\,\Gamma_{l,2l-1}^{\,
\infty,
\,\alpha_{0}}(0)=0\quad\mbox{for}\quad l>0\ ,
\end{equation}

we obtain in analogy to Lemma 1 
\vskip 0.3cm
{\bf Lemma 8.}
\begin{equation}\label{intr}
\partial_{p}^{\omega}\,i\,\Gamma^{\,\alpha,\,\alpha_{0}}_{l,s}(\vec
p)=\int\limits_{0}^{1}d\lambda_{1}\ldots\int\limits_{0}^{1}
d\lambda_{\sigma(l,s)}
\int\limits_{\alpha_{0}}^{\infty}d\alpha_{1}\ldots\int
\limits_{\alpha_{0}}^{\infty}d\alpha_{s}\
\partial_{p}^{\omega}\,G^{\,\alpha}_{l,s}(\vec \alpha,\vec \lambda,\vec p)
\quad,
\end{equation}
{\it and $\,\partial_{p}^{\omega}\,G_{l,s}^{\,\alpha}\,$ obeys the bounds}
\begin{equation}
\vert\partial_{p}^{\omega}\,G^{\,\alpha}_{l,s}(\vec\alpha,\vec\lambda,
\vec p)\vert\ \leq\
e^{\,-\varepsilon \,m^{2}\sum_{j=1}^{s}\alpha_{j}}\,P^{\,\varepsilon}
(\vert\vec p\vert)\,Q(\sqrt\alpha_{1},\ldots,\sqrt\alpha_{s})\quad,
\end{equation}
{\it where $\,P^{\,\varepsilon}\,$ is a polynomial with nonnegative 
coefficients
-- independent of $\,\alpha\,$ -- \\in $\,\vert p_{0,1}\vert,\ldots,
\vert p_{3,n-1}\vert\,$ and $\,Q\,$ is a nonnegative rational function 
which has no poles for $\,\alpha_{i}>0\, $.}
\vskip 0.3cm
{\it Proof.} We refer to the proof of Lemma 1 in section
4.1. $\,\Box\,$
\vskip 0.3cm
In order to examine the convergence of (\ref{intr}) as
$\,\alpha_{0}\rightarrow 0\,$ -- keeping $\,\varepsilon >0\,$ and the
external momenta real -- we now transfer our results of section 4.2
and state 
\vskip 0.3cm 
{\bf Proposition 9.} {\it The integral representation of the
$\,\partial_{p}^{\omega}\,i\,\Gamma_{l,s}^{\,\alpha,\,\alpha_{0}}\,$  
obeys the bounds}
\begin{displaymath}
\int\limits_{0}^{1}d\lambda_{1}
\ldots\int\limits_{0}^{1}d\lambda_{\sigma(l,s)}\int\limits_{
\alpha_{0}}^{\infty}d\alpha_{1}\ldots\int\limits_{\alpha_{0}}^{\infty}
d\alpha_{s}\,\vert\,\partial_{p}^{\omega}\,G_{l,s}^{\,\alpha}
(\vec \alpha,\vec
\lambda,p_{1},\ldots,p_{n-1})\,\vert
\end{displaymath}
\begin{equation}\label{th9}
\leq
\cases{\ P_{1}^{\,\varepsilon}(\,\vert\vec p\vert\,) & for 
$\ \alpha\geq\hat \alpha$ \cr
\ P_{2}^{\,\varepsilon}(\,\vert\log(\alpha)
\vert\,)\ P_{3}^{\,\varepsilon}(\,\vert\vec p\vert\sqrt\alpha\,)\ 
\alpha^{\,-2l+s+\frac{\vert\omega\vert}{2}} & for 
$\ \alpha_{0}\leq \alpha \leq
\hat \alpha\quad.$\cr }
\end{equation}
\vskip 0.3 cm
{\it $ P_{k}^{\,\varepsilon} \, $ are (each time they appear possibly new) 
polynomials with 
nonnegative coefficients which depend
neither on $\,\alpha\,$ nor on $\,\alpha_{0}\,$, but on
$\,\varepsilon,l,s,
\hat\alpha.\,$ $\,\hat \alpha>\alpha_{0}\,$ 
is some finite fixed number (e.g.1).} 
\vskip 0.3cm
{\it Proof.} See the proof of Proposition 2. 
$\,\Box\,$ 

From Proposition 9 we directly get
\vskip 0.3cm
{\bf Proposition 10.} 
{\it The one-particle irreducible renormalized Green 
functions of perturbative relativistic massive $\,\Phi_{\,4}^{\,4}\,$
with an $\,\varepsilon$-regularization given by (\ref{rpr}) and
(\ref{eta})}
\begin{equation}
i\,\Gamma_{l,s}^{\,\infty,\,0}(p_{1},\ldots,p_{n-1})
\end{equation}
{\it are well-defined for $\,\varepsilon>0\,$ and smooth with respect to 
$\,(p_{1},\ldots,
p_{n-1})\in{\rm\bf R}^{4(n-1)}\,$.
The integrands of their integral representations (\ref{intr}) are absolutely
integrable.}
\subsection{The Limit $\,\varepsilon\rightarrow 0\,$}
In order to analyse the limit $\,\varepsilon\rightarrow 0\,$ we need
more information about the structure of the integrands
$\,G_{l,s}^{\,\alpha}\,$. Using our results of section 5 we can state
\vskip 0.3cm
{\bf Proposition 11.} {\it The integrands $\,G_{l,s}^{\,\alpha}\,$ of the 
renormalized one-particle irreducible relativistic Green functions 
$\,i\,\Gamma_{l,s}^{\,\alpha,\,0}\,$ with an
$\,\varepsilon$-regularization given by (\ref{rpr}), (\ref{eta}) and
the renormalization conditions (\ref{renr}) have  the 
following structure:}

\begin{equation}\label{ga0r}
G^{\,\alpha}_{l,s}(\vec \alpha,\vec \lambda,\vec
p\,)\ =\ \sum_{j}\,P_{j}^{\,\varepsilon}(\,\vec p\,)
\,Q_{j}
(\vec \alpha,\vec
\lambda)\,e^{\,-\sum_{k,v}\,A^{\,j}_{kv}(\vec\alpha,\vec
\lambda)\, p_{k}\eta p_{v}}\,
V_{j}^{\,\alpha}(\vec \alpha)
\end{equation}
\begin{displaymath}
(\varepsilon-i)^{-\frac{1}{2}l}
(\varepsilon+i)^{-\frac{3}{2}l}
\,e^{\,-(\varepsilon+i)m^{2}\sum_{k=1}^{s}\alpha_{k}}\,
\end{displaymath}

{\it{\rm (a)} $\,A^{\,j}_{kv}(\vec \alpha,\vec \lambda)\, $, 
$\,Q_{j}(\vec\alpha,\vec\lambda)\,$ and
$\,V_{j}^{\,\alpha}(\vec\alpha)\,$ are  
identical to the corresponding functions in the 
Euclidean theory.

{\rm (b)} $\,P_{j}^{\,\varepsilon}(\vec p)=\prod_{\,k\leq v}\,(p_{k}
\,\eta\,p_{v})^{\,
u_{k,v}^{\,j}}\,$ (with the same $\,u_{k,v}^{\,j}\,$ as
in the Euclidean theory) are monomials, which for $\,\varepsilon=0\,$ are 
invariant under Lorentz transfor\-mations.}
\vskip 0.3cm
{\it Proof.} See the proof of Proposition 6.
$\,\Box\,$

In analogy to Corollary 7 we can easily prove
\vskip 0.3cm
{\bf Corollary 12.} {\it The integral representation (\ref{intr}) of the 
one-particle irreducible renormalized relativistic Green functions 
$\,i\,\Gamma_{l,s}^{\,
\infty,\,0}\,$ with an $\,\varepsilon$-regularization ((\ref{rpr}),
(\ref{eta})) and the renormalization conditions (\ref{renr}) can be written 
as}
\begin{displaymath}
i\,\Gamma^{\,\infty,\,0}_{l,s}(\,\vec
p\,)\,=\int\limits_{0}^{1}
d\vec\lambda\int\limits_{0}^{1}d\beta_{1}\ldots\int\limits_{0}^{1}
d\beta_{s}\,\delta\Bigl(1-\sum_{k=1}^{s}\beta_{k}\Bigr)\Biggl [\,\sum_{j}
\,V_{j}(\vec\beta\,)\,P_{j}^{\,\varepsilon}(\,\vec p\,)\,
Q_{j}(\vec\beta,\vec\lambda)
\end{displaymath}
\begin{equation}\label{pr7r}
\frac{(\varepsilon-i)^{-\frac{1}{2}l}\,(\varepsilon+i)^{-\frac{3}{2}l}
\,i^{\,d_{j}+s}}
{\Bigl (\,\sum_{k,v}A^{j}_{kv}(\vec\beta,\vec
\lambda)\,i\,p_{k}\eta p_{v}+\,(i\varepsilon-1)\,m^{2}\sum_{k=1}^{s}\beta_{k}
\Bigr )^{\,d_{j}+s}}
\ \Biggr ]\quad.
\end{equation}
\vskip 0.3cm
{\it The integrand is absolutely integrable. The momentum derivatives 
$\,\partial_{p}^{\omega}\,
i\,\Gamma_{l,s}^{\,\infty,\,0}\,$ are represented by integrals of the 
corresponding momentum derivatives of the integrand which are also absolutely
integrable.}
\vskip 0.3cm
{\it Proof.} See the proof of Corollary 7. $\,\Box\,$
\vskip 0.3cm
Let $\,\Psi(\vec p)\in S({\rm\bf R}^{4(n-1)})\,$ and
\begin{equation}\label{den}
F_{j}^{\,\varepsilon}(\Psi,\vec\beta,\vec\lambda)=\int d^{\,4}\vec p
\ \Psi(\vec p)\,\frac{P_{j}^{\,
\varepsilon}(\vec
p)\,(\varepsilon-i)^{\,-\frac{1}{2}l}\,(\varepsilon+i)^{\,-\frac{3}{2}l}}
{\Bigl (\,\sum_{k,v}A^{j}_{kv}(\vec\beta,\vec
\lambda)\,i\,p_{k}\eta p_{v}+\,(i\varepsilon-1)\,m^{2}\sum_{k=1}^{s}\beta_{k}
\Bigr )^{\,d_{j}+s}}\quad .
\end{equation}
Looking at the denominator in (\ref{den}) we realize that it has the
structure 
\begin{equation}\label{speer}
\left (\,{\rm P}_{1}\,+\,i\,\varepsilon\,{\rm
P}_{2}\,-\,m^{2}+i\,\varepsilon\,m^{2}\right)^{\,z} 
\end{equation}
with
\begin{equation}\label{dist1}
{\rm P}_{1}\ =\ \sum_{k,v}\,A^{\,j}_{kv}(\vec\beta,\vec \lambda)\bigl (
p_{0,k}p_{0,v}-\underline{p}_{k}\,\underline{p}_{v}\bigr)
\end{equation}
and 
\begin{equation}\label{dist2}
{\rm P}_{2}\ =\ \sum_{k,v}\,A^{\,j}_{kv}(\vec
\beta,\vec\lambda)\bigl(p_{0,k}p_{0,v}+\underline{p}_{k}\,
\underline{p}_{v}\bigr)\quad.
\end{equation}
Because $\,A^{\,j}(\vec\beta,\vec\lambda)\,$ is positive semi-definite
and continuous in $\,\vec\beta,\vec\lambda\,$ in the compact region
of integration we can apply a theorem due to Speer (p.105 \cite{spe}) that 
tells us that
for $\,\varepsilon\rightarrow 0 \,$ (\ref{speer}) defines a 
tempered
distribution which depends continuously on $\,A^{\,j}
(\vec\beta,\vec\lambda)\,$ and that (\ref{speer}) has the same distributional
limit as 
\begin{equation}
\left (\,{\rm P}_{1}\,-\,m^{2}+i\,\varepsilon\,m^{2}\right)^{\,z} 
\end{equation}
which is Lorentz invariant. Therefore we conclude that
for $\,\varepsilon\rightarrow 0 \,$ (\ref{den}) defines a 
Lorentz invariant tempered distribution which depends continuously on 
$\,A^{\,j}(\vec\beta,\vec\lambda)\,$.

In the compact region of integration                                    
\begin{equation}\label{eps0}
\sum_{j}\,V_{j}(\vec\beta\,)\,Q_{j}(\vec\beta,\vec\lambda)\,F_{j}^{\,
\varepsilon}(\Psi,\vec\beta,\vec\lambda)
\end{equation}
is absolutely integrable for all $\,\varepsilon>0\,$ and $\,F_{j}^{\,0}
(\Psi,\vec\beta,\vec\lambda)\,$ is continuous. 
Therefore we conclude that (\ref{eps0}) is still absolutely
integrable for $\,\varepsilon=0\,$.

Thus we can state
\vskip 0.3cm
{\bf Theorem 13.} 
{\it The limit $\,\varepsilon\rightarrow 0\,$ of 
the one-particle irreducible 
renormalized Green 
functions of perturbative relativistic massive $\,\Phi_{\,4}^{\,4}\,$
defined by (\ref{rpr}), (\ref{eta}), (\ref{lre}) and (\ref{renr})}
\begin{equation}
\lim_{\varepsilon\rightarrow 0}\ i\,\Gamma_{l,s}^{\,\infty,\,0}(p_{1},
\ldots,p_{n-1})
\end{equation}
{\it exists as a Lorentz invariant tempered distribution 
$\,\in S'(\,{\rm\bf R}^{4(n-1)})\,$.}
\vskip 0.3cm
{\it Remark.} Using the flow equation (\ref{rfl}) written for the 
perturbative,
regularized amputated connected Green functions $\,i{\cal
L}_{l,s}^{\,\alpha,\,\alpha_{0}}(\vec p)\,$ a theorem analogous
to Theorem 13 for these Green functions could be proved by a
similar line of argumentation.
\vskip 0.3cm
Comparing the corresponding integrands of the integral representations in the
(multiplied by $\,(-1)\,$)
Euclidean theory (\ref{pr7})
and in the relativistic theory
(\ref{pr7r}) we realize that they coincide as functions up to a factor
$\,(-i)^{s-l}\,$ if  
in (\ref{pr7}) the imaginary parts $\,k_{0,v}\,$ of the zero 
components of the external momenta take values in the domain
(\ref{img}) and the real parts are equal to 0,
and if we let $\,\varepsilon\rightarrow 0\,$ and set $\,p_{0,v}=
k_{0,v}\,$ in (\ref{pr7r}).

According to Theorem 5 these integrands are absolutely integrable and 
therefore we can conclude
\vskip 0.3cm
{\bf Theorem 14.} 
{\it The limit $\,\varepsilon\rightarrow 0\,$ of 
the one-particle irreducible 
renormalized Green 
functions of perturbative relativistic massive $\,\Phi_{\,4}^{\,4}\,$
defined by (\ref{rpr}), (\ref{eta}),  (\ref{lre}) and (\ref{renr})}
\begin{equation}
\lim_{\varepsilon\rightarrow 0}\ i\,\Gamma_{l,s}^{\,\infty,\,0}(p_{1},
\ldots,p_{n-1})
\end{equation}
{\it exist as Lorentz invariant smooth functions in the domain $D'$ defined
as follows:}
\begin{displaymath}
D:=\biggl\{(p_{0,1},\ldots,p_{0,n-1})\in {\rm\bf R}^{n-1}
\mbox{ {\it with} }\ 
\vert\,\sum_{j\in\tau_{a}}p_{0,j}\,\vert<2m\quad\mbox{  {\it for all} 
}\quad
\tau_{a}\subseteq\{1,\ldots,n\}\ ,
\end{displaymath}
{\it $\,
p_{0,n}=-\sum_{j=1}^{n-1}p_{0,j}\,$ and $\,
(\underline{p}_{1},\ldots,
\underline{p}_{n-1})\in{\rm\bf R}^{3(n-1)}\biggr\}\quad ;\quad D':=
\bigcup_{\Lambda\in L}\,\Lambda\, 
D\quad,\ L\cong$ Lorentz group.} 

\end{document}